\documentclass[useAMS,usenatbib]{mn2e}
\usepackage[hidelinks]{hyperref}
\usepackage{graphicx} 
\usepackage{times} 
\newcommand{\seq}{\,{=}\,}

\defcitealias{petersen18a}{Paper I}
\defcitealias{petersen18b}{Paper II}
\defcitealias{petersen18e}{Paper V}
\defcitealias{petersen18d}{Paper IV}

\def\kms{ {\rm km~s\textsuperscript{-1}}}
\def\Lz{  L_{\rm z}}

\begin{document}

\title[Barred Galaxy Harmonic Decomposition]
 {Using Harmonic Decomposition to Understand Barred Galaxy Evolution}

\author[Petersen, Weinberg, \& Katz] 
{Michael~S.~Petersen,$^1$\thanks{mpete0@astro.umass.edu} Martin~D.~Weinberg,$^1$ Neal~Katz$^1$ \\ 
$^1$University of Massachusetts at Amherst, 710 N. Pleasant St., Amherst, MA 01003}

 \maketitle 
\begin{abstract} 
We study the mechanisms and evolutionary phases of bar formation in $n$-body simulations of a stellar disc and dark matter halo system using harmonic basis function expansion analysis to characterize the dynamical mechanisms in bar evolution. We correlate orbit families with phases of bar evolution by using empirical orthogonal functions that act as a spatial filter and form the gravitational potential basis. In both models we find evidence for three phases in evolution with unique harmonic signatures. We recover known analytic results, such as bar slowdown owing to angular momentum transfer. We also find new dynamical mechanisms for bar evolution: a steady-state equilibrium configuration and harmonic interaction resulting in harmonic mode locking, both of which may be observable. Additionally, we find that ellipse fitting may severely overestimate measurements of bar length by a factor of two relative to the measurements based on orbits that comprise the true backbone supporting the bar feature. The bias will lead to overestimates of both bar mass and bar pattern speed, affecting inferences about the evolution of bars in the real universe, such as the fraction of bars with fast pattern speeds. We propose a direct observational technique to compute the radial extent of trapped orbits and determine a dynamical length for the bar.
\end{abstract} 


\section{Introduction} \label{sec:introduction} 

The clear presence of responses to non-axisymmetric disturbances in galaxies---bars, spiral arms, warps, rings, and displacements, amongst other features---necessitate a higher-order harmonic description of stellar discs beyond an exponentially-decreasing monopole. Early studies characterized disc structure using Fourier amplitudes in rings of  radius $R$, $A_m(R) \seq\frac{1}{2\pi}\int_{-\pi}^{\pi} f(R,\phi) e^{-im\phi}d\phi$ where $m$ is the harmonic order and $f(R,\phi)$ is the weighting function corresponding to the luminosity (or ideally mass) as a function of azimuthal angle $\phi$ around the galaxy \citep{considere88, elmegreen89}. Barred galaxies, which make up more than half of the observed disc galaxies in the infrared \citep{sheth08}, are the most pronounced examples of galaxies with large values of the quadrupole $A_2(\equiv\int A_m(R)dR)$ in the inner galaxy. Owing to the ease with which it is computed, $A_2$ has long been used as a proxy for the `strength' of a bar--a nebulously defined term that does not fully or necessarily accurately quantify the effect of the bar on the evolution of the galaxy, as we will illustrate below.

One can move beyond Fourier analysis in rings. Performing a basis function expansion (BFE) that correlates spatial and azimuthal structure and more accurately represents the gravitational field that causes the non-axisymmetric structures. One may then use the BFE measures to study the evolutionary mechanisms and scenarios for evolution. Connecting dynamical principles to galaxy evolutionary mechanisms to the BFE allows one to better  study the evolutionary phases of barred galaxy evolution. In addition, a harmonic BFE analysis is an inexpensive way to parameterize both the evolution of large simulations and observational data. 

The harmonic BFE decomposition technique has been used to both study and compare simulations, owing to its natural relationship with analytic perturbation theory \citep{weinberg07a,weinberg07b}. Some $n$-body simulations use a technique explicitly built on biorthogonal functions, where one solves the Poisson equation using separable azimuthal harmonics. Generally, these techniques may be called BFE  \citep{cluttonbrock72, cluttonbrock73, kalnajs76, hernquist92, earn96, weinberg99}, which has many notable features that make them ideal for studying disturbances to equilibrium stellar discs. For simulations using BFE methods, harmonic function analysis decomposes a distribution into linearly-summable functions that resemble expected evolutionary scenarios in disc galaxy evolution. The primary diagnostics available are the amplitude and phase of each function. When tracked through time, one unlocks another dimension for studying evolution that may not clearly manifest itself in analytic studies \citep{weinberg04}. Using harmonic function analysis in BFE simulations enables a quick and straightforward reconstruction of the potential at any time in the simulation, for any arbitrary combination of particles. Reconstructions of arbitrary potentials has already allowed us to locate resonances using perturbation theory and commensurability mapping \cite[][hereafter Paper I]{petersen18a} and to determine which components and channels are responsible for the primary transfer of angular momentum \cite[][hereafter Paper II]{petersen18b}. In one of the applications of BFE in this work, we determine that many measurements of the bar, in particular the bar length, may be biased by large-scale structure in the galaxy, leading to overestimates for the mass of the bar and the pattern speed.

The goal of this paper is analyse the evolutionary mechanisms in a stellar disc and determine the evolutionary phases of a barred galaxy model using a BFE method. To build a dynamical picture from evolutionary scenarios, we seek answers to the following questions: (1) Which mechanisms dominate the evolution? (2) What observables do the mechanisms have? (3) How do mechanisms interact with one another? By defining functions that describe evolutionary scenarios and mechanisms, the BFE method enables us to readily identify features correlated by self-gravity. We then use the BFE functions to study the dynamical mechanisms responsible for the evolution. In this paper, we show that a wealth of responses can result from secular evolution alone, provided that the phase-space admits channels for secular evolution. Some of our results match previous findings, such as the slowdown of the bar, but many describe new dynamics, including harmonic-mode-locking as a mechanism to slow or stop bar evolution. Harmonic mode locking is a newly-discovered mechanism that may connect real, observed features of galaxies to underlying dynamics. We find that harmonic mode locking can halt the evolution of a model barred galaxy.

The paper is organized as follows. In Section~\ref{sec:methods}, we motivate the choice of our BFE methodology and describe its details, present the simulations, and summarise our analysis performed in previous work (Papers I and II) that we employ throughout this paper. In Section~\ref{sec:barmodes}, we look at the global measurements of the simulation using the harmonics from the basis, including a direct comparison of ellipse fitting and dynamical measurements from harmonic function analysis. In Section~\ref{sec:reconstruction}, we quantify the bar feature itself using BFE, studying both azimuthal and radial representations of the bar. In Section~\ref{sec:discussion}, we discuss harmonics relating directly to the bar and their amplitudes versus time to help identify mechanisms and evolutionary phases. Some harmonics and their related mechanisms are either not present in our simulation as expected, or present and appear to have less influence than expected. Section~\ref{sec:observations} presents a kinematic technique with which we determine the true length of orbits that form the backbone of the bar. We conclude in Section~\ref{sec:conclusion}.

\section{Methods} \label{sec:methods}

We compute the potential as the sum of two orthonormal basis sets describing the equilibrium of a galaxy disc (the first basis) and the dark matter halo (the second basis). Our primary tool is the BFE method as implemented in {\sc exp} \citep{weinberg99}. The BFE method has four primary advantages over tree and grid method gravity solvers: (1) the calculation of forces scales linearly with particle number, (2) the dynamic range of multi-scale systems such as the disc-halo system can be better resolved by tailoring the geometry and scale of the components individually, (3) a sensitivity to weak global distortions is possible because small-scale noise can be removed, and (4) intercomponent interactions can be explicitly controlled to allow the study of different mechanisms individually. The self-gravity and intercomponent forces between all components can be independently selected and controlled. We discuss the details of our implementation of the BFE algorithm in Section~\ref{subsec:eof}, the initial conditions of our simulations in Section~\ref{subsec:initialconditions}, and our methods for measuring the bar in Section~\ref{subsec:barmeasurement}.

\subsection{Empirical Orthogonal Functions} \label{subsec:eof}

\begin{figure*} \centering \includegraphics[width=6.5in]{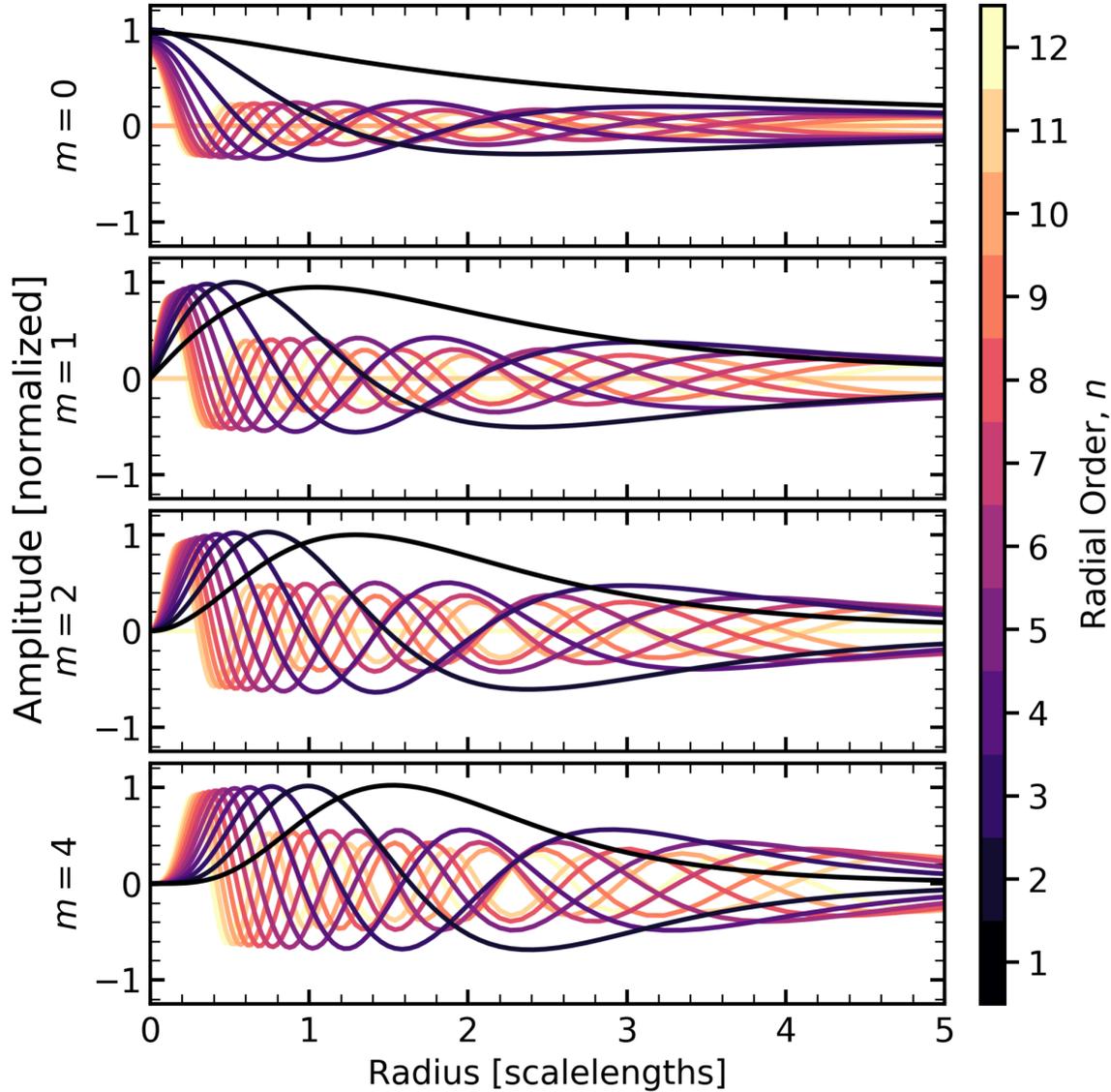} 
\caption{\label{fig:disc_amplitudes} In-plane amplitude variations as a function of disc scalelength for all radial functions per harmonic order in the cylindrical disc basis. We show the $m\seq0,1,2,4$ harmonic subspaces as panels from top to bottom. The amplitude in each panel has been normalized to the maximum in the corresponding subspace. Functions that are zero everywhere are vertically asymmetric (see Figure~\protect{\ref{fig:disc_3d_amplitudes}}).} \end{figure*}

\begin{figure} \centering \includegraphics[width=3.5in]{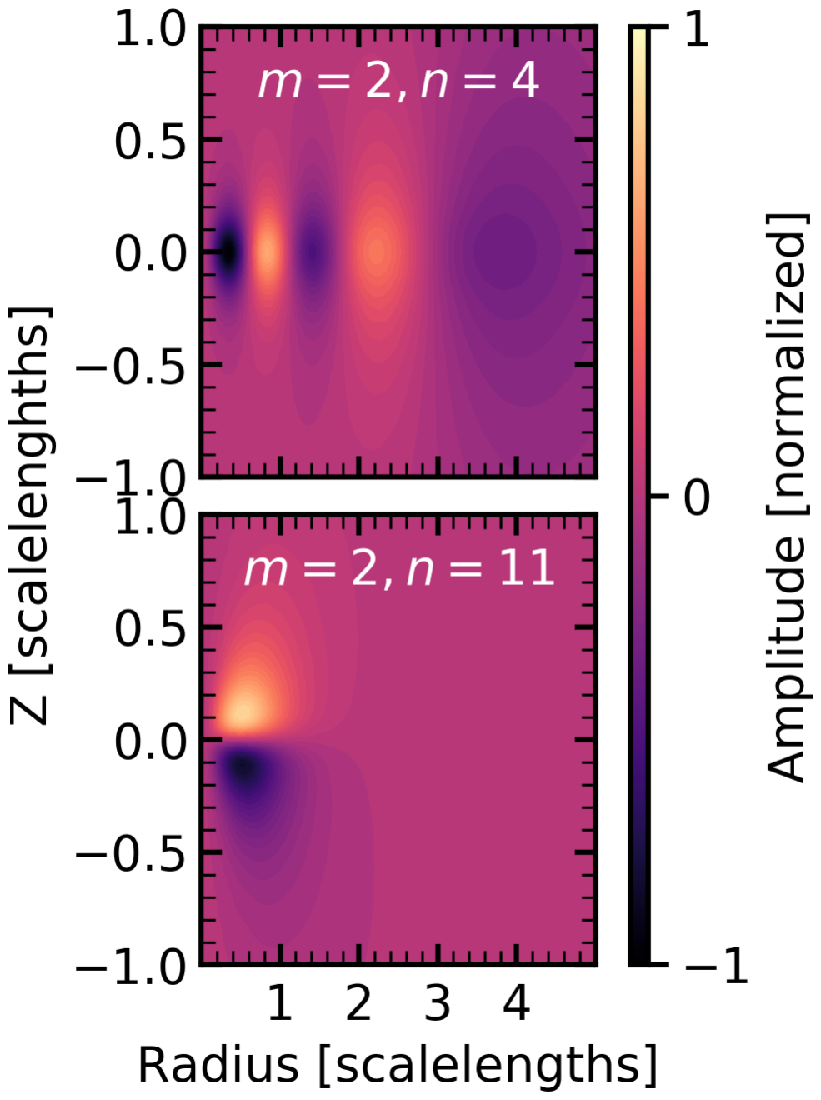} 
\caption{\label{fig:disc_3d_amplitudes} Examples of vertically symmetric ($m\seq2,n\seq4$, upper panel), and vertically asymmetric ($m\seq2,n\seq11$, lower panel) functions for the disc basis. The $x$ and $z$ axis correspond to the radial and vertical axes in the simulation, and the amplitude of the variations between panels has been normalized to the maximum $m\seq2$ amplitude. } \end{figure}

In the BFE method one computes the gravitational potential by projecting particles onto a set of biorthogonal basis functions that satisfy the Poisson equation. One then evaluates the force term for each particle at the position of each particle. This approach relies on the properties of solutions to the Sturm-Louiville equation (SLE) of which the Poisson equation is a special case. The SLE describes many physical systems, and may be written as: 
\begin{equation}
\frac{d}{dx}\left[p(x)\frac{d\Phi(x)}{dx}\right] - q(x)\Phi(x) \seq \lambda \omega(x) \Phi(x)
\label{eq:sle}
\end{equation}
where $\lambda$ is a constant, and $\omega(x)$ is a weighting function. The eigenfunctions $\phi_j$ of the SLE form a complete basis set with eigenfunctions $\lambda_j$, where $j$ may be truncated from the theoretically infinite series. When applied to the Poisson equation specifically, the Fourier and Bessel expansions are two well-known examples. The BFE potential solver is built using properties of eigenfunctions and eigenvalues of the SLE. 

The halo can be expanded into a relatively small number of spherical harmonics $Y_{lm}$ and appropriate radial functions, such that the total halo potential is given by $\Phi_{lm}^j \seq \phi_{lm}^j(r)Y_{lm}(\theta,\phi)$. The disc is more complicated, and requires a specially constructed basis. As the Poisson equation may be written as an eigenfunction of the Laplacian, which has solutions that are a product of spherical harmonics in the angular variables and Bessel functions in radius \citep{weinberg99}, solutions to the SLE may be reduced to a separable form in cylindrical coordinates $r$,$z$, and $\theta$ with radial, vertical, and azimuthal functions\footnote{The Poisson equation can be separated in any conic system; the choice of cylindrical coordinates is motivated by the geometry of the disc.} that satisfy a potential of the form $\Phi(r) \seq R(r)Z(z)\Theta(\theta)$ with a corresponding density function. 

Although one can construct a disc basis from the eigenfunctions of the Laplacian, the boundary conditions make the basis hard to implement. To get around this, our solution method starts with a spherical basis with $l\le36$ and uses a singular value decomposition to define a rotation in function space to best represent a target disc density. The new basis functions optimally approximate the true distribution in a linear least squares sense. The new eigenfunctions are also orthonormal and representable as a linear combination of solutions to the Poisson equation. Therefore, the new functions are also solutions to the Poisson equation. Because we are free to break up the spherical basis into meriodinal subspaces by azimuthal order, the resulting two-dimensional eigenfunctions in $r$ and $\theta$ are equivalent to a decomposition in cylindrical coordinates $r,~z,$ and $\theta$. These techniques have been packaged into the $n$-body code {\sc exp}, which we use for our simulations.

We condition the initial disc basis functions on the analytic disc density such that the lowest-order potential-density pair matches the initial analytic mass distribution, parameterized in our simulation as in equation~\ref{eq:exponentialdisc}. This acts to reduce small-scale discreteness noise as compared to conditioning the basis function on the realized positions of the particles \citep{weinberg98}, although there could be some other biases introduced by this procedure. Both simulations presented in this paper use the same disc basis, allowing for a detailed comparison between differences in the disc profiles. Throughout this paper, we refer to disc azimuthal harmonics as $m$--orders, and radial subspaces as $n$--orders, such that an eigenfunction is given by $(m,n)$ notation. In the halo, the azimuthal harmonics are $l$ orders, with $m\in(-l,l)$, as in spherical harmonics. Radial subspaces in the halo are still referred to as $n$--orders. In this paper, we focus on the disc harmonics to motivate our discussion of observational harmonic decomposition of barred disc galaxies. 

Using basis methods, we can understand the potential and density of a galaxy as a superposition of several basis functions. This allows us to decompose the galaxy into harmonic orders based on their symmetry, where $m\seq0$ is the monopole, $m\seq1$ is the dipole, $m\seq2$ is the quadrupole, and so on. The sine and cosine terms of each azimuthal order give the phase angle of the harmonic that can be used to calculate the pattern speed. We also decompose the azimuthal harmonics into radial subspaces that also set the vertical structure.  After trial-and-error, we determined that a radial scale factor for the spherical profile of approximately $\sqrt{2}$ was appropriate for setting the scale of the disc when deprojected onto spherical harmonics for computation of the basis. As we shall see later, this choice of radial scale does not appear to offset the radial subspaces that exhibit significant amplitude. 

Figure~\ref{fig:disc_amplitudes} shows the in-plane amplitude variations for radial functions ($n$ orders) as a function of radius, separated by harmonic subspace ($m$ orders). We show the four harmonic subspaces that are most relevant for the evolution of the simulation, $m\seq0,1,2,4$, from top to bottom in the panels. In each harmonic subspace, the lowest-order radial order, $n\seq1$, has no nodes except at $R=0$ for $m\ge0$. The number of nodes increases with order $n$. The nodes are interleaved by radial order, but the increasing number of nodes means that the smallest radius node always decreases in radius as the number of nodes increases. Therefore, an increase in amplitude for higher--$n$--order harmonics corresponds to the movement of mass to smaller radii. Additionally, the spacing of nodes gives an approximate value for the force resolution of the simulation. For example, the highest order $m\seq0$ radial function ($n\seq12$) has a zero at $R\seq0.2a$, or 600 pc in a MW-like galaxy. Additionally, the radial orders are interleaved between harmonic orders, such that $R_{\rm first node,m\seq2,n\seq1} \approx \frac{1}{2}\left(R_{\rm first node,m\seq1,n\seq1} + R_{\rm first node,m\seq1,n\seq2}\right)$. $n_{\rm max}$ is selected to provide a spatial resolution of $\approx$500 pc. Remember, however, that the lowest-order basis function exactly matches the initial density profile. For example, in the halo basis even though the highest $n$ order would only imply a spatial resolution of 100 pc, the basis resolves a power law in density down to 10 pc. This choice removes or filters high spatial frequencies that may increase relaxation noise. In Figure~\ref{fig:disc_3d_amplitudes}, we show examples of the vertical structure in the disc basis functions. The upper panel shows the $m\seq2,n\seq4$ basis function in radius--z space. This function is symmetric about the $z\seq0$ axis. The combination of vertically symmetric and asymmetric harmonics represent all possible variations in the gravitational field above and below the plane consistent with the spatial scales in the basis. In both panels, the color has been normalized to the maximum amplitude of the $m\seq2$ harmonic subspace. 

For each eigenfunction, we sum over the particle distribution and compute the contribution to the coefficients at each timestep in the simulation. We define our potential estimator in one dimension as
\begin{equation}
a_j = \int dx\bar{\phi}_j(x)f(x)
\end{equation}
where $\int dx \bar{\phi}_j(x)\phi_k(x) \seq \delta_{jk}$ satisfies the orthogonality relation and $f(x)$ is some well-defined function, in this case the true potential \citep{weinberg96}. In the case of tabulated eigenfunctions and a discrete distribution of particles, such as in an $n$-body simulation, the coefficients that approximate the potential are
\begin{equation}
\hat{a}_j = \frac{1}{N} \sum_{k=1}^N \phi_j(x_k)
\end{equation}
where $\phi_j$ is the potential eigenfunction that satisfies the biorthogonality relation, as above. Then, if we take $\hat{a}_j$ to be an estimate for $a_j$, we can estimate the function $f(x)$, here representative of the potential, as
\begin{equation}
\hat{f}(x) = \sum_{j=1}^M \hat{a}_j\phi_j(x).
\end{equation}

Throughout this work, we will evaluate and report $\hat{a}_j$, which we will refer to as a `coefficient' or `amplitude' of a particular eigenfunction. Tracking the amplitudes for the basis functions through time is the primary investigative tool used in this paper. The coefficients for each $n$ order have cosine and sine components that correspond to the analogous Fourier terms $A_m$ and $B_m$. Thus we may compute the phase angle for any basis function. When we combine the sine and cosine terms to make a single amplitude or modulus for the particular $(m,n)$ eigenfunction, we will use the notation $A_{m,n}$. The total amplitude in an azimuthal harmonic order will simply be noted as $A_m$. 

The BFE approach also has tradeoffs. The truncated series of basis functions intentionally limits the possible degrees of freedom in the gravitational field; one must investigate whether the basis can capture all possible mechanisms of disc evolution. However, a basis function representation provides an information--rich summary of the gravitational field and provides insight into the overall evolution. As we will see in Section~\ref{subsec:barcoefficients}, this method allows for the decomposition of different components into dynamically-relevant subcomponents for which the gravitational field can be calculated separately. For brevity and instructive comparison to previous work, we will refer to the harmonic decomposition employed here as Fourier, but we emphasize that the decomposition of galaxy models using orthogonal functions goes beyond traditional Fourier techniques. For example, our BFE method accounts for both the radial harmonics and vertical structure of the disc galaxy, with few assumptions beyond the initial conditions.

\subsection{Initial Conditions}\label{subsec:initialconditions}

This paper focuses on the detailed results from two simulations with initial conditions that illustrate the differences between evolution in cuspy and cored dark matter halos. We briefly describe the initial conditions and refer the reader to previous papers that introduced the simulations in more depth (\citetalias{petersen18a} and \citetalias{petersen18b}). The initial condition procedure is also discussed in \citet{holleyb05} and \citet{petersen16a}. We adopt $G\seq1$ and virial units where $M_{\rm vir}\seq R_{\rm vir}\seq1$, $V_{\rm vir}\seq1$, $T_{\rm vir}\seq1$. The simulations may be scaled to obtain physical quantities for different galaxies. Appropriate scalings for the Milky Way are $M_{\rm vir}=1.4\times10^{12}$ M$_{\odot}$, $R_{\rm vir}=300$ kpc, $V_{\rm vir}=220$ \kms, and $T_{\rm vir}=2$ Gyr.

The simulations begin with an exponential disc with density
\begin{equation}
\rho_{\rm disc}(R,z) = \frac{M_{\rm d}}{8\pi ha^2} e^{-R/a} {\rm sech}^2 (z/h)
\label{eq:exponentialdisc}
\end{equation}
where $M_d\seq0.025M_{\rm vir}$ is the disc mass, $a\seq0.01R_{\rm vir}$ is the disc scale length, and $h\seq0.001R_{\rm vir}$ is the disc scale height. We embed the disc in a modified NFW \citep{navarro97} dark matter halo with $c\seq25$, whose density is given by
\begin{equation}
\rho_{\rm halo}(R) = \frac{\rho_0r_s^3}{\left(R+r_c\right)\left(R+r_s\right)^2}
\label{eq:nfw}
\end{equation}
where $\rho_0$ is a normalization set by the chosen mass, $r_s\seq R_{\rm vir}/c$ is the scale radius for the turnover of the exponents, and $r_c$ is a radius that sets the size of a core. The core radius $r_c$ distinguishes between the two models: $r_c\seq0$ in the cusp simulation, and $r_c\seq0.02$ in the core simulation. The adjustable core radius allows us to explore the role of halo profile on secular dynamics.

We realize the initial positions and velocities in the halo via Eddington inversion of the halo model that includes the monopole contribution from the disc. We select the initial positions in the disc (equation~\ref{eq:exponentialdisc}) via an acceptance--rejection algorithm. We select the velocities by solving the Jeans equations with an axisymmetric velocity ellipsoid in the disc plane ($\sigma_r\seq\sigma_\phi$). We characterize the velocity dispersion using the Toomre $Q$ parameter,
\begin{equation}
\sigma_r^2(r) = \frac{3.36\Sigma(r)Q}{\Omega_r(r)}
\end{equation}
where $\Sigma(r)$ is the surface density and $\Omega_r$, the radial frequency, is given by
\begin{equation}
\Omega_r^2(r) = r\frac{d\Omega_\phi^2}{dR}+4\Omega_\phi^2.
\end{equation}
where $\Omega_\phi$ is the azimuthal frequency. We choose $Q\seq0.9$, a `cold' disc, to promote the rapid growth of disc structure. The vertical velocity dispersion is obtained directly from the Jeans' equations for a disc,
\begin{equation}
\sigma_z^2(r) = \frac{1}{\rho_d(R,z)}\int_z^\infty \rho_d(R,z)\frac{\partial\Phi_{\rm tot}}{\partial z}dz
\end{equation}
where $\Phi_{\rm tot}$ is the sum of the disc and halo potential \citep{binney08}. 

As both simulations use the same basis for the disc, we can compare the excited basis amplitudes directly between the two simulations. Although BFE reproduces any potential field in principle, truncation of the series limits its full adaptability. 

\subsection{Bar Measurement} \label{subsec:barmeasurement}

We use two methods to parameterize the size and mass of the bar: traditional ellipse fitting to the isophotes of surface density and measuring the radii of the trapped orbits that support the bar potential.

\subsubsection{Ellipse Fits}\label{subsubsec:ellipsefits}

Many studies have made use of visually-determined bar lengths, including Galaxy Zoo \citep{hoyle11} and S$^4$G \citep{sheth08}. While visually measuring a bar length is a quick process, the method doesn't necessarily trace an isodensity surface and often offers no errors on individual measurements.  \citet{hoyle11} found that individual observers report approximately a 6 per cent deviation relative to the mean of all observers who classify a bar length. 

Other studies fit ellipses to bar isophotes \citep{munoz13, laurikainen14, kim15, erwin16, kruk18}. The various ellipse measurements have known discrepancies. \citet{athanassoula02} demonstrate that different ellipse methods applied to the same galaxy can lead to variations of up to 35 per cent in measured bar length. However, as this method is commonly used, we also adopt it here. We fit isophotes using least-squares regression to a standard ellipse equation. Some studies use a generalized ellipse \citep{athanassoula90}, where the ellipse equation is given by
\begin{equation}
1 = \left(\frac{|x|}{a}\right)^c + \left(\frac{|y|}{b}\right)^c.
\label{eq:genellipse}
\end{equation}
The standard conic ellipse assumes $c\seq2$, while the generalized ellipse allows for a variable $c$. Values $c>2$ yield `boxy' isophotes. As pointed out by \cite{athanassoula13}, the bar length can be overestimated relative to visual classification when $c$ is not allowed to vary. In this work, we do not allow $c$ to vary, and acknowledge that some fits may result in longer bars than the values reported here. However, we find that our results do not qualitatively change if we use a generalized ellipse instead. In our tests, the relative variance in the length of the fit ellipse was approximately 25 per cent if $c$ is a fit parameter\footnote{Clearly, allowing $c$ to vary changes the amount of $m\seq4$ amplitude in the bar feature, which has discernable dynamical consequences \protect{\citepalias{petersen18a}}.}.

\citet{munoz13} measured bar lengths in the S$^4$G sample \citep{sheth08} using four different ellipse metrics derived from either the ellipticity profile or position angle of the best-fit ellipse at a given radius. Connecting these methods, \cite{herreraendoqui15} demonstrated that visual fits to bar lengths are roughly comparable to lengths determined from the radius of maximum ellipticity. We compute the best fit ellipses over a range of isophotal values in the vicintity of the bar, and we assign the bar length as the maximum semi-major axis that has $b/a<0.5$. We find that selecting different criteria such as the maximum ellipticity or a threshold in position angle variation does not qualitatively change our results; all commonly-used ellipse measures return values within approximately 30 per cent, in agreement with \citet{athanassoula02}. For a more thorough introduction to measuring bar lengths, we refer the reader to \citet{erwin05} for an observationally motivated viewpoint, and \citet{athanassoula02} for a theoretically motivated viewpoint.

\subsubsection{Trapping Analysis} \label{subsubsec:trappedanalysis}

In \citetalias{petersen18a}, we analyzed bar membership through the clustering of the radial turning points, or apsides, for a given orbit. Orbits `trapped' by the bar's gravity will librate about the position angle of the bar's major axis. In this work, we primarily consider orbits that contribute to the structure of the bar: the $x_1$ family associated with the inner Lindblad resonance (ILR; $2\Omega_\phi-\Omega_r \seq 2\Omega_p$, where $\Omega_p$ is the pattern frequency of the bar). These orbits comprise the `backbone' of the bar and are eccentric orbits elongated along the bar axis. We also consider bar supporting orbits that are composed of higher-order families that reinforce the bar potential but are not directly associated with ILR; we denote these as `other'.

Briefly described, our method isolates the turning points in an orbit by looking for local maxima in radius. Using a rolling average of 20 apsides\footnote{We determinte the rolling average of 20 empirically to be a sweet spot in a tradeoff between time resolution and signal-to-noise.} in Cartesian coordinates that we transform to a frame co-rotating with the bar position, $\theta_{\rm bar}$, we compute the position angle for the center of two $k$--means--derived clusters relative to the bar, taking the maximum of the two values. The choice of two clusters is motivated to align with the two ends of the bar, as a trapped orbit will librate around the minimum of the potential caused by the bar, analogous to a swinging pendulum librating around its mimumum. In addition to the cluster position angles, we compute the variance in the position angle relative to the cluster center over the 20 apsides, $\sigma_{\theta_{\rm bar,20}}$. These two quantities alone allow for a successful classification of orbits into the $x_1$ and `other' bar supporting families as follows. We limit the average apse position to be $\langle\theta_{\rm bar}\rangle \le \pi/8$ then subdivide based on the variance in $x_1$ with $\sigma_{\theta_{\rm bar,20}}\le \pi/16$ and `other' with $\pi/16 < \sigma_{\theta_{\rm bar,20}} \le 3\pi/16$. From an empirical examination of the orbits, we estimate a contamination rate in both families of approximately 1 per cent. This uncertainty does not change any of the results we present in this work.

\section{A BFE-based View of Bar Phases} \label{sec:barmodes}

The BFE potential--solving methodology of \textsc{exp} naturally lends itself to harmonic function analysis. We present the harmonic decomposition of the simulations, using both the amplitude and phase of basis function coefficients to characterize the global evolution of the simulations. Then we may determine distinct evolutionary phases and  apply complementary analyses such as perturbation theory, orbit analysis, and torque theory. In \citetalias{petersen18a}, we first identified three phases of bar evolution from the trapped fraction of orbits: assembly, growth, and steady state. For a fixed basis, the simulation results may be efficiently summarised and compared using the time series of coefficients. Here, we will correlate the phases of bar evolution identified from the trapped fraction with the harmonic decomposition of the simulation phase space. We first describe the evolution of azimuthal harmonics and introduce the three phases of bar evolution observed in our simulations in Section~\ref{subsec:azimuthal} before considering the information in the radial subspaces in Section~\ref{subsec:radialsubspaces}. We summarize the utility of harmonic decomposition for a detailed study of the evolution in Section~\ref{subsec:modesummary}.

\subsection{Azimuthal Harmonics} \label{subsec:azimuthal}

\begin{figure*} \centering \includegraphics[width=6.5in]{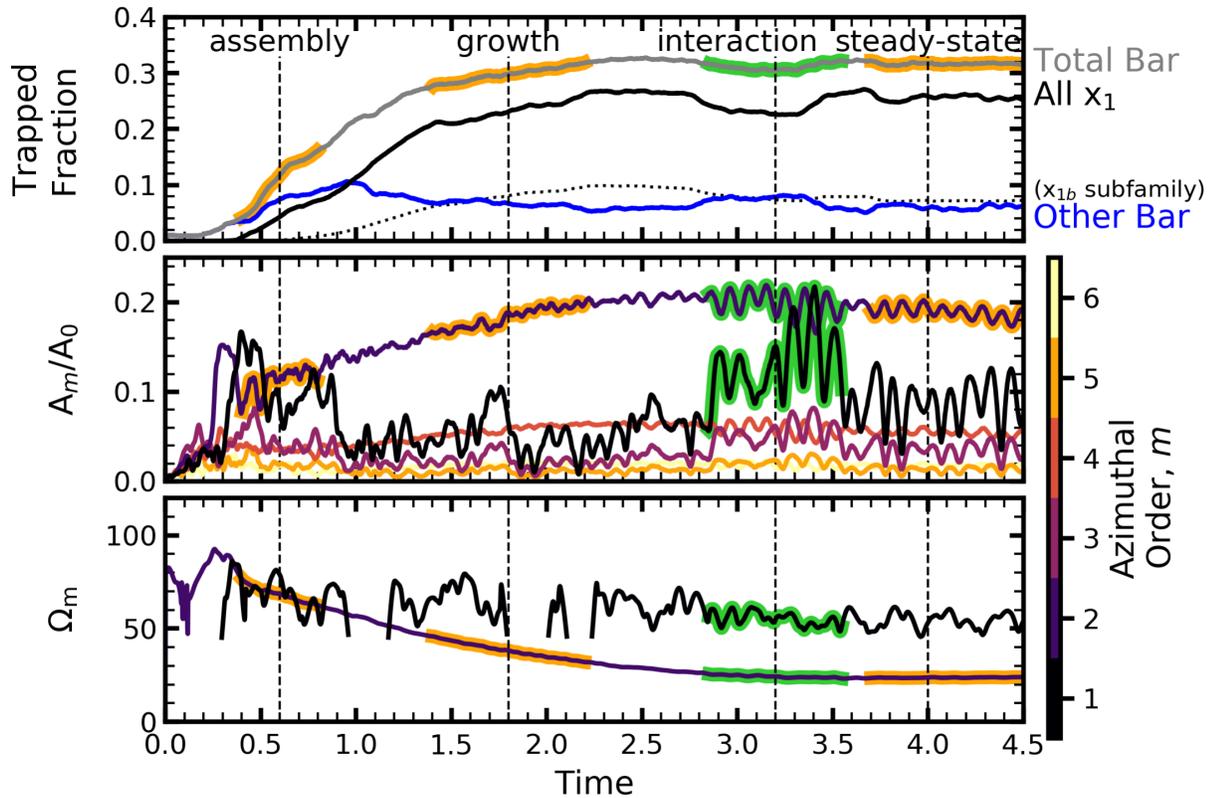} 
\caption{\label{fig:cusp_phases} Upper panel: Bar membership in the primary $x_1$ (black) and `other' bar-supporting (blue) families, and total (gray), vs. time, from the cusp simulation. Middle panel: $A_m$, the amplitude per harmonic orders $m\in(1,6)$. Lower panel: azimuthal harmonic pattern speed, $\Omega_{m}$, where $m\in(1,2)$. Windows of low phase signal are not plotted. The evolution of higher harmonic orders of $m$ in pattern speed are the same as the other odd or even harmonics, respectively. In all panels, three prominent phases in the bar lifetime are identified and highlighted in orange: {\it assembly}, {\it growth}, and {\it steady-state}. A prominent {\it interaction} between the $m\seq1$ and $m\seq2$ harmonics is highlighted in green. } \end{figure*}

\begin{figure*} \centering \includegraphics[width=6.5in]{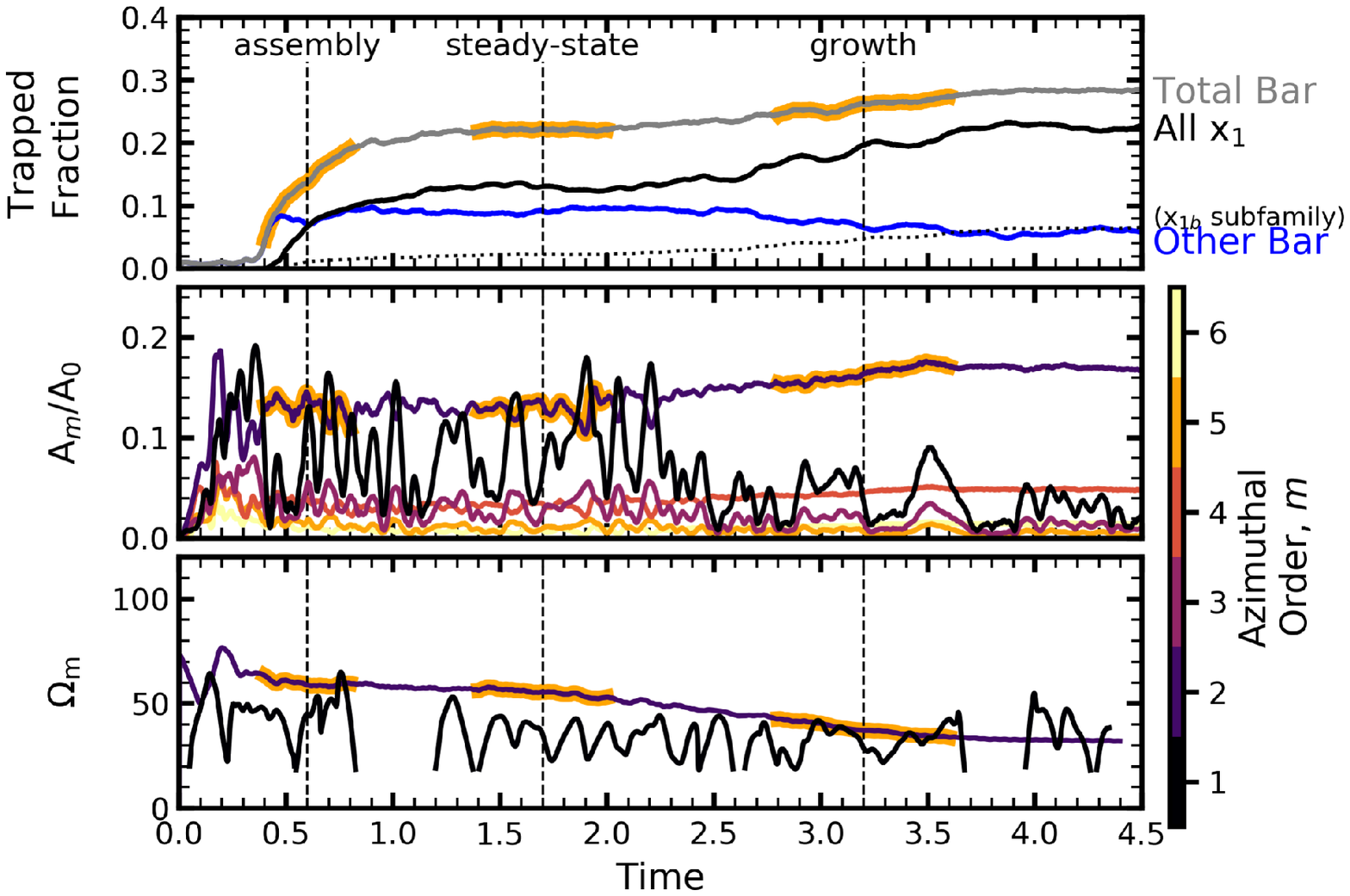} 
\caption{\label{fig:core_phases} Same as Figure~\protect{\ref{fig:cusp_phases}}, but for the core simulation. In all panels, three prominent phases in the bar lifetime are identified and highlighted, in order: {\it assembly}, {\it steady-state}, and {\it growth}. } \end{figure*}

With the trapped fraction analysis of \citetalias{petersen18a} in hand, we can use our a priori knowledge about the basis and the qualitative features of the evolution to derive a diagnostic classification in terms of the basis. Figures~\ref{fig:cusp_phases} and \ref{fig:core_phases} show the three phases of bar evolution \citepalias[described in][]{petersen18a}, the corresponding coefficient power, and the pattern speed derived from the coefficients for the cusp and core simulations. In particular, we wish to draw contrasts between the cusp and core simulation, showing that a mild change to the halo profile can produce evolution that is qualitatively different.

The upper panel of Figure~\ref{fig:cusp_phases} plots the trapped orbit fraction versus time for the total bar (gray), and the two orbit families that make up the bar, the $x_1$ and higher-order (`other') bar supporting families (black and blue, respectively). We also show a bifurcated subfamily of the $x_1$ orbits, the $x_{1b}$ family, as a dotted black line.

We identify and label three qualitative phases of bar evolution in orange: the assembly phase, the secular growth phase, and the steady-state equilibrium state. We characterize the phases as follows: (1) the assembly phase, where the trapped fraction grows quickly (assembly); (2) a secular growth phase where the bar continues to growth linearly (growth); and (3) a steady-state phase where the trapped fraction do not increase, but other global quantities may change \citepalias[see][for additional discussion]{petersen18b}. The evolution of the trapped fraction distinguishes between the phases while a visual, i.e. isophotal, inspection of the bars does not \citepalias{petersen18b}. The middle panel of Figure~\ref{fig:cusp_phases} shows the total contribution to each azimuthal harmonic order computed by summing over the radial orders for fixed $m\in(1,6)$. The colors correspond to the harmonic order as indicated. Any single azimuthal harmonic is not clearly associated with the growth of distinct orbit subfamilies, though $m=2$ is correlated with the growth of the total bar. However, we can see signs of the three qualitative phases of evolution.

As expected, the trends in $m\seq2$ azimuthal power correlate with the total trapped fraction and with each of the three observed phases. The trapped bar is the `true' bar from a dynamical standpoint; identifying the trapped component is key for dynamical interpretation. The assembly phase lags the $m\seq2$ feature that is traditionally associated in the literature with the formation of the bar. 

A rapid increase and then decrease of the $m\seq2$ amplitude at $T\seq0.3$ is clearly associated with significant transfer of angular momentum to the outer disc via the two-armed spiral that precedes the formation of the trapped bar \citepalias[see][]{petersen18b}\footnote{A lower disc-to-halo mass ratio or a warm disc leads to a less significant initial rearrangement.}. An examination of the radial coefficients for $m\seq2$ reveals that the spiral arms and not the bar are responsible for the initial growth of the $A_2$ component. Unfortunately, the trapped component is difficult to distinguish observationally, though we offer a possible method in Section~\ref{sec:observations}. Whether measured by bar mass or $m\seq2$ amplitude, the bar strength grows consistently over time. However, when measured by Fourier amplitude, the bar appears to assemble quickly and strengthen more slowly than the trapped orbits indicate. 

In the cusp simulation, the overall value of the amplitude of the $m\seq2$ harmonic subspace differs from the trapped fraction by 50 per cent during the phases where the bar is clearly established (growth and steady-state). This cautions against using Fourier techniques alone to quantify the strength of the bar, particularly during the assembly phase. Further, the $m\seq1$ amplitude is greater than that of $m\seq2$ at two key points during the simulation: the assembly phase and also at the harmonic--coupling phase, a nonlinear power transfer between two harmonics, $m\seq1$ and $m\seq2$ at $T\approx3$. We will analyze the harmonic subspace coupling in much more detail in later sections. Apart from these two times, the $m\seq1$ harmonic subspace is subdominant in amplitude, often lower than the $m\seq4$ amplitude. As with the even harmonics, the higher-order odd harmonics ($m\seq3,5$) qualitatively resemble the evolution of the $m\seq1$ harmonic.

To check for any dependence on the basis center, we perform one additional simulation to study the effects of excluding $m\seq1$ harmonics from barred galaxy evolution. In this simulation, which uses the same initial conditions as the cusp simulation, we do not allow forces for any odd harmonics to be applied to the particles. We refer to this simulation as the `even--harmonic--only' cusp simulation. Analysis of the even--harmonic--only cusp simulation, where we artificially enforce $\hat{a}_{m} \seq 0$ for all radial subspaces $m\in\{1,3,5\}$ suggests that: (1) the $m\seq1$ harmonic subspace is important for the formation of the bar, and (2) the $m\seq1$ harmonic subspace is necessary for the long-term stability of the bar, particularly as it grows. In the even--harmonic--only cusp simulation, the bar that forms is only 75 per cent as strong as the bar in the fiducial simulation, despite having identical initial conditions.

The bottom panel of Figure~\ref{fig:cusp_phases} shows the pattern speed for the $m\seq1$ ($\Omega_1$) and $m\seq2$ ($\Omega_2$) harmonic orders, color--coded as in the middle panel. The higher--order harmonic orders exhibit the same pattern speed as the $m\seq1$ and $m\seq2$ orders for the higher odd and even harmonic orders, respectively. One can interpret the $m\seq2$ pattern as the pattern speed of the bar ($\Omega_p$), particularly during the growth and steady-state phases. Outer disc activity moderately affects the assembly phase, evident in the mismatch between the relatively large $m\seq2$ amplitude and the relatively small trapped fraction during the assembly phase as previously discussed. The $m\seq1$ pattern passes through the center for much of the simulation (up until $T\seq3$), exhibiting a radial `sloshing' or seiche mode where the phase angle of the $m\seq1$ amplitude becomes zero. Near $T\seq3$, the $m\seq1$ pattern becomes locked to the phase of $m\seq2$, and begins rotating with the bar rather than oscillating radially. We highlight this time in green and label this as the `interaction' phase. The interaction and locking of the harmonics reveals a new mechanism for the bar to transfer angular momentum: the bar pattern transfers power to the $m\seq1$ pattern, imposing net rotation on the seiche mode, and causing the entire bar to orbit the center of mass of the combined disc-halo system.

The core simulation behaves similarly (compare Figure~\ref{fig:core_phases} to Figure~\ref{fig:cusp_phases}). However, as discussed in Papers I and II, the onset of the growth phase occurs after an extended, nearly steady-state phase. For this reason, the evolutionary phases in the core simulation proceed as assembly, steady-state, and finally growth, as shown in the top panel of Figure~\ref{fig:core_phases}. The trapping of a bifurcation of the $x_1$ family, the $x_{1b}$ subfamily, dominates the growth phase. The ratio of $A_4/A_2$ increases with time just prior to and during the growth phase, which signals the appearance the $x_1$ bifurcation that drives the growth phase \citepalias[see][]{petersen18a}. In the core model, the $m\seq1$ amplitude has comparable maximum values to that of the $m\seq2$ harmonic subspace during both the assembly and steady-state phases. However, the $m=1$ amplitude is significantly more volatile in the core simulation. 

The bottom panel of Figure~\ref{fig:core_phases} shows the pattern speed of the $m\seq2$ and $m\seq1$ harmonics, color--coded as in the middle panel. The $m\seq2$ pattern speed evolves as expected from standard secular-evolution theory. During the growth phase, the pattern speed slows significantly and during the steady-state phase, the pattern speed remains roughly constant, with a small decrease. Harmonic deomposition is unlikely to be informative for bar evolution during the assembly phase, as in the cusp simulation, owing to the strong contribution by the outer disc to the $m\seq2$ amplitude. In contrast to the cusp simulation, the $m\seq1$ pattern begins at a much lower pattern speed and begins as a seiche mode, passing back and forth through the center. During the steady-state phase, the $m\seq1$ pattern rotates about the center, as in the cusp simulation.

In contrast to many previous studies, we emphasize and characterize the different phases of bar evolution. The largest difference is that the bar stops slowing. We have not identified the precise mechanisms that halt the bar slowing. However, it is intriguing that in both models the asymptotic value of the $m\seq2$ pattern speed ($\Omega_2$) is some low integer fraction of the $m\seq1$ pattern speed, $\Omega_1$ ($\frac{1}{2}$ in the cusp model, 1 in the core model). In the cusp model, it appears that the $m\seq2$ harmonic subspace pattern speed approaches a low-integer commensurability with the non-evolving  pattern speed of the $m\seq1$ harmonic subspace. This commensurability breaks the secular evolution channel by causing $\Omega_2$ to oscillate, inhibiting resonance passage. In the cusp model $\Omega_2$ asymptotes to $0.5\Omega_1$, which means that the bar is displaced from the center in a rotating pattern such that when the bar completes a half rotation the $m\seq1$ pattern has completed a full rotation. Visually, this appears as a constant offset of the bar from the center of the galaxy. The maximum displacement of the center is $0.3a$, which would be approximately 1 kpc in the MW. In the core model $\Omega_2$ approaches $\Omega_1$, but there is no obvious phase locking. We discuss the implications of these processes for studies of galaxy evolution in Section~\ref{subsubsec:m1}.

\subsection{Radial Subspaces} \label{subsec:radialsubspaces}

\begin{figure*} \centering \includegraphics[width=6.5in]{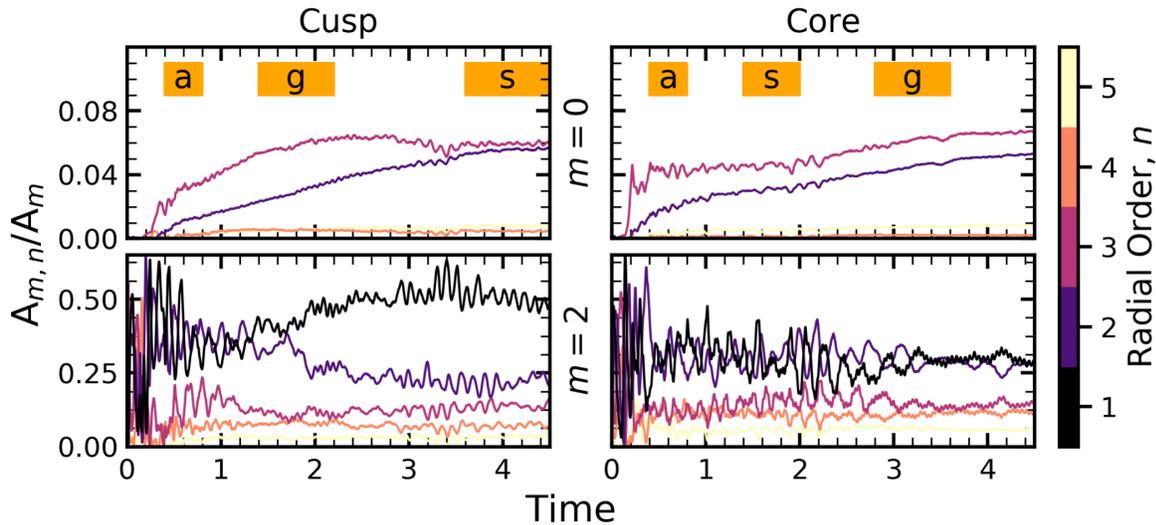} 
\caption{\label{fig:radial_order} Amplitude in low-order radial subspaces ($n$) of the $m\seq0$ and $m\seq2$ azimuthal harmonic subspaces, normalized by the total amplitude in the corresponding harmonic subspace. We show $n\in(2,5)$ for the $m=0$ subspace, and $n\in(1,5)$ for the $m=2$ subspace. The left panels are for the  cusp simulation, and the right panels are for the core simulation. The upper row is $m\seq0$ and the bottom row is $m\seq2$.} \end{figure*}

For any azimuthal subspace $m$, the amplitude of the radial harmonics indicate the radial scale of a feature or response, as described in Section~\ref{subsec:eof}. Radial and azimuthal orders together provide more detailed information than total azimuthal power, e.g. $A_2$. Remember that as before, we use the nomenclature $(m,n)$ to refer to the specific radial subspaces $n$ of a given harmonic order $m$. For example, the $m\seq2,n\seq2$ function will be denoted by $(2,2)$. When we discuss an entire harmonic subspace, for example the $m\seq2$ harmonic subspace, we denote that as $(2,n)$. 

Consider, for example, the monopole $(0,n)$. The $m\seq0$ subspace has no phase information, and the amplitude is simply set by the total mass in the model (which we use as a normalization in the previous section). However, the partitioning of the $m\seq0$ harmonic subspace into ranges by radial order can reveal the scale of any changes in the galaxy model. In Figure~\ref{fig:radial_order}, we show the first five radial orders for the $m\seq0$ subspace (top row) and $m\seq2$ subspace (bottom row). The left column corresponds to the cusp simulation, and the right column corresponds to the core simulation. In each row, the individual lines have been normalized by the total amplitude in the corresponding harmonic subspace, e.g. for the $m\seq2,n\seq2$ amplitude we plot $A_{m,n}/A_m$, using the notation described in Section~\ref{subsec:eof}.

The $(0,0)$ amplitude always dominates as one would expect by construction since the lowest-order basis function exactly matches the initial model\footnote{Owing to the large amplitude of the $(0,0)$ function, we do not show it in Figure~\ref{fig:radial_order}.}. The total amplitude of the $(0,n>0)$ radial subspaces does not exceed 13 per cent during the entire integration. Additionally, throughout the simulation, the ordering of the radial harmonics largely stays intact (i.e. $A_{m,n}>A_{m,n+1}$ for all $n$ orders) for the $m>0$ harmonic subspaces. Further, the radial orders are all in phase, except for the harmonic mode locking that occurs at $T\seq3.0$ in the cusp simulation, where the radial orders can be misaligned by up to $\pi/4$. Considering the $(0,n)$ radial subspaces for both simulations (the upper row of Figure~\ref{fig:radial_order}), we see that the initial radial ordering does not apply. For $m\seq0$, the $n\seq2$ radial harmonic ($R_{\rm first node} \seq 0.6a$) describes the rearrangement of the disk mass when the bar forms. The $(0,1)$ subspace ($R_{\rm first node} \seq 1.2a$) grows nearly linearly in both simulations, suggesting a gradual rearrangement of the disc by the presence of the bar. The higher order [$(0,n>2)$, $R_{\rm first node} < 0.4a$] harmonics of the $m\seq0$ subspace play a subdominant role, never exceeding more than 1 per cent {\it in total} of the $m\seq0$ amplitude, suggesting that there is little small-scale rearrangement of the disc, and that the evolution is driven by the lowest-order radial harmonics.

The $m\seq2$ harmonic subspace, which is responsible for the bar feature, may also be decomposed into radial orders. Here we see stark differences between the cusp and core simulations. Comparing the radial decomposition of the $m\seq2$ order to the overall $m\seq2$ amplitude in Figure~\ref{fig:cusp_phases} for the cusp simulation, we see that the assembly phase consists of equal parts $(2,0)$ and $(2,1)$ amplitude. The growth phase results in the $(2,0)$ amplitude increasing while the $(2,1)$ amplitude decreases in relative importance, before all $n$ orders more-or-less stop evolving by the steady-state phase\footnote{All $n$ orders appear to participate in the harmonic-locking at $T\approx3$, exhibiting higher variance over the interaction phase described above.}. The increase in the $(2,0)$ amplitude is consistent with the lengthening of the bar, such that the order with the largest node spacing will gain proportionally more amplitude. The bar eventually becomes longer than $R_{\rm first node}$ of the $(2,1)$ term, which suggests that the bar length cannot be described solely by a single radial term; using the nodes specifically to understand bar shape is subtle. In Section~\ref{subsec:barreconstruction}, we study reconstruction of the bar from the radial orders of the density functions. 

The $n>4$ orders combine to have less than 5 per cent of the total $m\seq2$ amplitude. In contrast, the core simulation reveals a bar composed of a significantly different distribution in the radial subspaces. At all times, the $(2,0)$ and $(2,1)$ amplitude are comparable, even as the bar lengthens throughout the simulation. Further, the amplitudes of the $(2,n>2)$ harmonics are larger than the analogous harmonics in the cusp simulation. The bar is thus supported by a wider spectrum of harmonics in the cored simulation than in the cusp simulation, a result of the different bar geometries. We cite this as evidence for the steady evolution of the bar in the cusp simulation, whereas the evolution bar in the core simulation is punctuated by periods of transformation, owing to the high variance in the $m\seq2$ radial orders destabilizing the evolution and prohibiting continued bar growth. When the variance in the radial harmonics of $m\seq2$ decreases at $T\approx3$ in the core simulation (lower right panel of Figure~\ref{fig:radial_order}), the bar then begins its secular growth phase.

\subsection{Summary} \label{subsec:modesummary}

Our key interpretations are as follows:
\begin{enumerate}
\item The $m\seq2$ total amplitude correlates with the evolution in fully-formed bars, i.e., the $m\seq2$ total amplitude traces bar growth in a bar-dominated galaxy, but is mixed with strong spiral arm activity when it exists. This is also true for the pattern speed; the total pattern speed of $m\seq2$ is the bar pattern speed at late times, but during assembly it may be biased by large-scale $m\seq2$ arm activity in the model.
\item The $m\seq1$ total amplitude plays a dynamically important role, and its presence should not be ignored. It both captures formation scenarios (as evidenced by the instability of the even--harmonic--only simulation comparison), and slows the evolution in the cusp simulation through harmonic interaction (discussed further below).
\item The $m\seq0,n\seq2$ function appears to have a special correspondence with the bar. As the first node of this function is near the length of the bar, this function may represent the demarcation of angular momentum by the bar length. Thus, a function with similar node spacing may be directly associated with the bar monopole in simulations generally.
\end{enumerate}
These interpretations are generally true between the two models, and provide a set of diagnostics with which to interpret additional simulations. We will make use of these results in future work.

\section{BFE Representations of Simulated Bars} \label{sec:reconstruction}

We use the basic findings from the previous section to offer a more nuanced look at the coefficients that make up the bar and how we can use these harmonics to learn about bar evolution. In Section~\ref{subsec:barcoefficients}, we extract the coefficients that pertain to the bar only both as a means of verification of the orbit methodology, and also as a means to examine harmonic function analysis in more depth. In Section~\ref{subsec:barreconstruction}, we use the BFE coefficients and functions to reconstruct the bar density. We learn how the structure of the bar is represented in the basis and the utility of parameterizing a bar with a BFE representation.

\subsection{Bar Coefficients}  \label{subsec:barcoefficients}

\begin{figure*} \centering \includegraphics[width=6.5in]{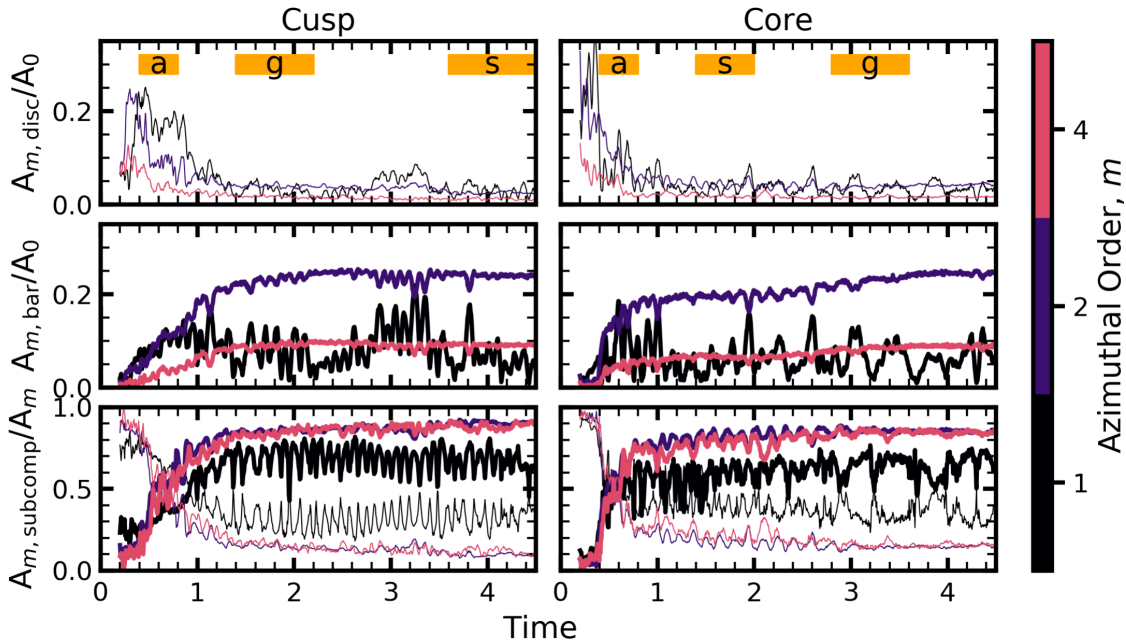} 
\caption{\label{fig:bar_fraction} Decomposition of power in azimuthal harmonics by contribution to the bar feature. The left column corresponds to the cusp simulation and the right column corresponds to the core simulation. Upper panels: the amplitude per harmonic order, for the untrapped disc only, normalized by the amplitude in the monopole. Middle panels: the amplitude per harmonic order, for the bar only, normalized by the amplitude in the monopole. Lower panels: the amplitude per harmonic order, subdivided by bar membership (thick solid lines) or non--bar membership (thin solid lines), normalized by the amplitude in the corresponding harmonic order. We show only the three strongest azimuthal orders by power normalized to the monopole, $m\in[1,2,4]$. The evolutionary phases in each column are labeled: assembly (a), growth (g), and steady-state (s).} \end{figure*}

In Section~\ref{sec:barmodes}, we described the correlations between the bar trapped fraction and the directly measured coefficient amplitudes. We use the trapped fraction to partially accumulate the coefficients for the bar. We then use this information to examine the evolution of the bar at each of the three phases. In Figure~\ref{fig:bar_fraction}, we decompose the three strongest azimuthal harmonic orders ($m\seq1,2,4$) into coefficients accumulated from particles not in the bar (upper row, thin dashed lines), coefficients accumulated from trapped bar particles (middle row, thick solid lines), and the combination (lowest row, line thickness as in upper two panels). The upper two rows of panels show the coefficients normalized to the monopole amplitude, and the lowest row of panels show the coefficients normalized to the total amplitude in the corresponding azimuthal harmonic order. The left column represents the cusp simulation and the right column the core simulation.

As expected, the $m\seq2$ amplitude for particles in the cusp simulation trapped in the bar strongly correlate with the trapped fraction (cf. Figure~\ref{fig:cusp_phases}), growing rapidly during the assembly phase (a), linearly growing during the growth phase (g), and remaining constant during the steady-state phase (s). The $m\seq2$ amplitude in the core simulation behaves similarly during its respective assembly (a), steady-state (s), and growth (g) phases. In both simulations, the $m\seq4$ components of the bar strongly resembles those of $m\seq2$. As discussed in \citetalias{petersen18a}, the $A_4/A_2$ ratio does not remain constant in either simulation. The $A_4/A_2$ ratio increases strongly during the growth phase for the core simulation in particular. Particles that are part of the bar dominate $A_2$ and $A_4$ for all the distinct evolutionary phases. $A_1$ is strongly affected by the bar with the exception of during the assembly phase in the cusp simulation, where $A_1$ is attributable to the untrapped disc particles rather than to the bar particles. It is also particularly evident in the upper panels that $m\seq1$ trades amplitude with the $m\seq2$ harmonic, and to a lesser extent with the $m\seq4$ harmonic. The initial burst of $m\seq2$ and $m\seq1$ power in the simulations is a rearrangement that results from initial conditions rather than the formation of the bar, which occurs at a more modest pace. The bar increases rapidly in strength until $T\approx0.5$, in contrast to the initial peak in total $m\seq2$ power that occurs at $T\approx0.1$.

The lower row of panels of Figure~\ref{fig:bar_fraction} shows the relative amplitude for bar and non-bar particles, which confirms that the bar is the dominant source of amplitude in all harmonic orders (though not shown, this is true for $m\seq3,5,6$ as well). At the end of both simulations, the bar accounts for greater than 80 per cent of the $m\seq2,4$ coefficient amplitude. Not only does this confirm that the bar itself is the major source of non-axisymmetric disc distortion, it also serves as an implicit check of the orbit determination method, described in \citetalias{petersen18a}. Orbits may be efficiently and unambiguously attributed to the bar and we are not missing any significant population of bar-supporting orbits. The importance of the bar in $m\seq1$ amplitude relative to the untrapped disc is surprising, but a subsequent orbit analysis shows that the bifurcated $x_1$ family, the $x_{1b}$ family, is asymmetric with respect to the center and can sustain an $m\seq1$ disturbance \citepalias{petersen18a}. This same family is responsible for the strong `beating' as the bar pattern speed approaches the natural $m\seq1$ frequency (see Section~\ref{subsubsec:m1}). It appears that the bar controls the majority of the $m\seq1$ amplitude at all times after assembly, in both simulations. 

\subsection{Bar Reconstruction}  \label{subsec:barreconstruction}

\begin{figure*} \centering \includegraphics[width=6.5in]{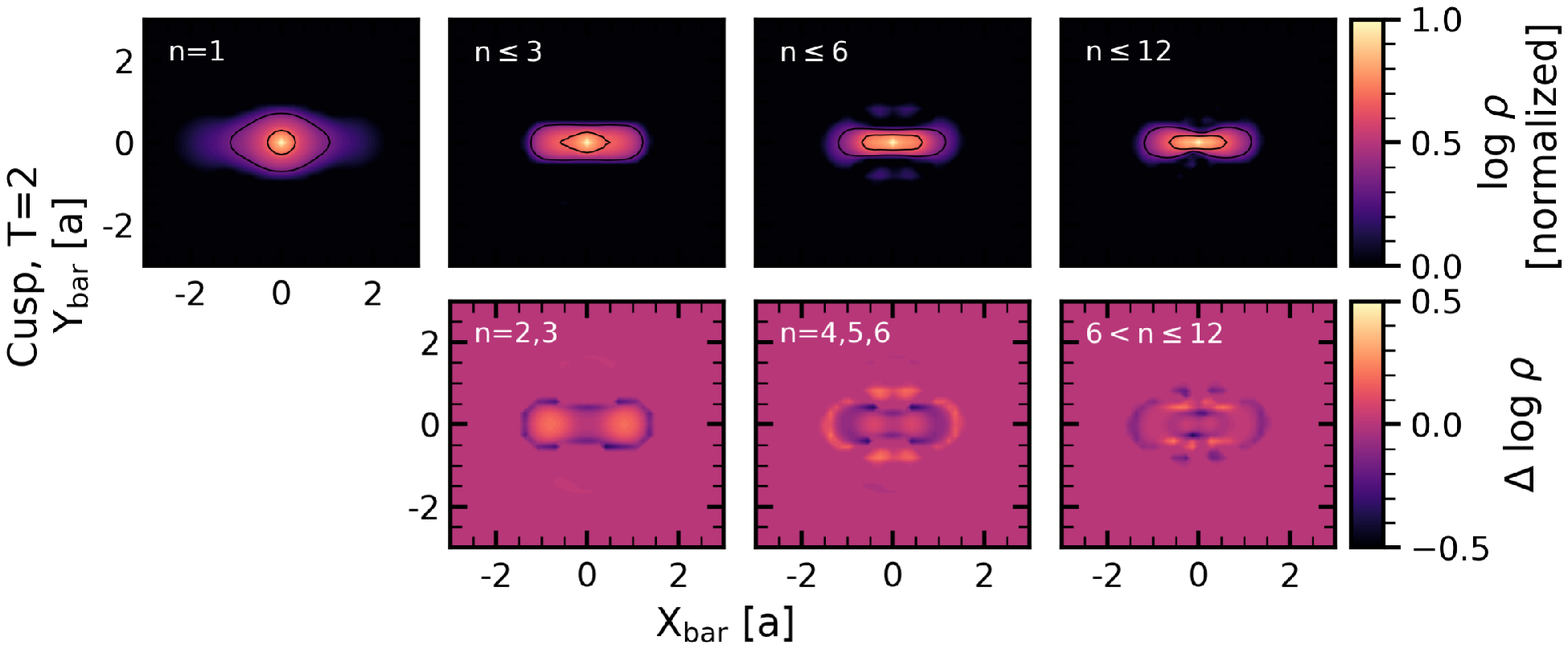} 
\caption{\label{fig:cusp_reconstruction} Reconstruction of the bar at $T=2$ in the cusp simulation. The upper row of panels shows the density of the bar as reconstructed from the functions. From left to right, we show the reconstruction of the bar using only the lowest radial order function ($n=1$) for each $m>0$ azimuthal order, then using the lowest three ($n\le3$), the lowest six ($n\le6$), and all the radial functions ($n\le12$). In each panel, we include all radial functions for the $m=0$ azimuthal harmonic so as to resolve the axisymmetric monopole structure. The non-axisymmetric structure of the bar is fully resolved by using all radial functions.The lower row of panels shows the contribution to the density by the intermediate radial orders---the difference between successive panels in the upper row. From left to right, the relative density from the $n=2,3$ terms, the $n=4,5,6$ terms, and the $n\in(7,12)$ terms.} \end{figure*}

\begin{figure*} \centering \includegraphics[width=6.5in]{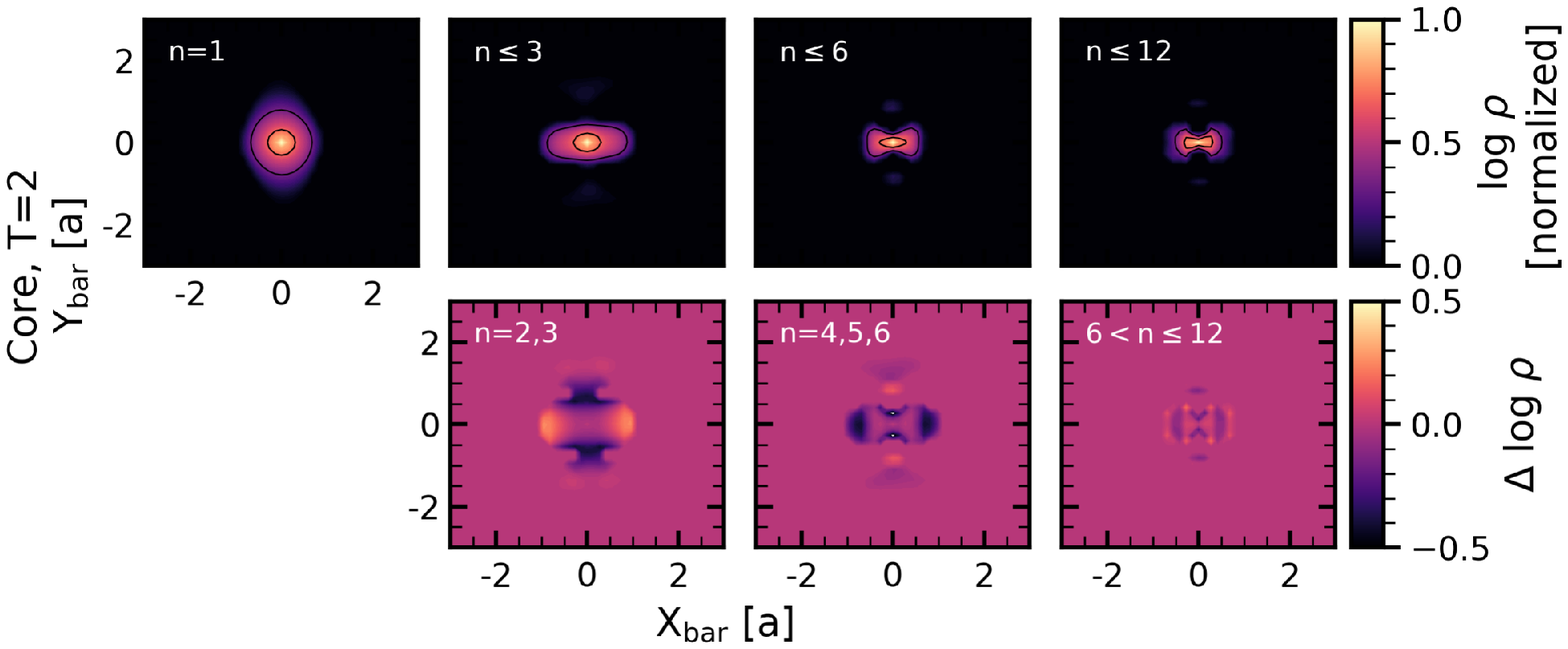} 
\caption{\label{fig:core_reconstruction} Same as Figure~\ref{fig:cusp_reconstruction}, except for the core simulation. In this representation of the bar, the lowest order non-axisymmetric radial terms ($m>0$, $n=1$) are oriented perpendicular to the bar.} \end{figure*}

The density functions of the BFE can be used to create reconstructions of the bar using partially accumuated coefficients from the bar particles. One may not only examine the entire azimuthal order as in Section~\ref{subsec:barcoefficients}, but may also study the structure of the bar as a function of radial orders. The bar reconstruction depends on the included number radial orders: the lower-order radial functions represent the gross structure of the bar, while the higher-order radial functions bring the fine structure of the bar into focus. We use the reconstruction as a sanity check to verify our choice of basis.

In Figures~\ref{fig:cusp_reconstruction} and \ref{fig:core_reconstruction}, we demonstrate the reconstruction of the bar by adding successive radial orders for the cusp and core simulations respectively at $T=2$. The upper rows show the density reconstruction using the partially accumulated coefficients and the density functions. We show the effect of adding successive radial orders by limiting the maximum radial order included from left to right, such that the leftmost panel shows only the lowest-order radial function and the rightmost panel is the complete reconstruction of the bar. As discussed above, the cusp and core simulations have the same disc basis. Comparing the $n\le12$ panels of Figures~\ref{fig:cusp_reconstruction} and \ref{fig:core_reconstruction} demonstrates the different geometries that may be represented by a single disc basis. As shown in Figure~\ref{fig:radial_order}, the coefficients corresponding to different $m=2$ radial orders are markedly different between the cusp and core simulation. Furthermore, the lowest-order radial function in the core simulation is oriented perpendicular to the bar to resolve the $m=2$ structure in the outer disc.

The lower rows of Figures~\ref{fig:cusp_reconstruction} and \ref{fig:core_reconstruction} demonstrate the finer structure of the bar achieved by adding successive radial orders. In the cusp simulation, the bar length is established with the inclusion of $n=3$, but the `waist' of the bar along the axis perpendicular to the bar is not resolved until $n=10$. In the core simulation, the bar length is not established until $n=6$, at which time the total structure of the bar is established.

While we have only shown one reconstruction of the bar in time, one may also study the changing structure of the bar through time by looking at the change in coefficients and the resulting reconstructions. The relative weights of the radial orders inform the structure of the bar, with longer bars having a higher proportion of the total azimuthal power in the lowest radial orders. We can, therefore, use the radial orders in the simulation to inform the structure of the bar as it evolves and to make a robust parameterization of bar structure through time for fixed potential applications.

\section{Dynamical Analysis} \label{sec:discussion}

We describe two modes present in the simulations and their importance for galactic evolution. In Section~\ref{subsec:verticalmodes}, we discuss the lack of buckling observed in our simulations, including physical and numerical reasons for why a vertical instability may not be present. In Section~\ref{subsec:planarmodes}, we look at two effects that are not typically included in linear analyses of potentials and discuss their importance for galaxy evolution: dipole ($m=1$) modes, which may control the late-time evolutionary state of the bar, and radial power exchange, which is a result of resolvable dynamical processes.

\subsection{Vertical Modes} \label{subsec:verticalmodes}

We begin with a brief review of the low order ($m\seq0,1$) bending modes. The lowest-order mode for a disc embedded in a halo is the sloshing or `seiche' mode \citep[e.g.][]{weinberg91}. Subsequent works divided this response into multiple classes: \cite{sparke95} presents a `bowl-shaped' $m\seq0$ bending mode that results from the disc sloshing through the halo midplane, flexing into a bowl shape\footnote{This may also be the `banana' mode that is sometimes used to describe polar ring galaxies.}. This mode is neutrally stable, meaning that it neither grows nor decays, in the case of a displacement from the vertical midplane. \cite{merritt94} describe a `bell' $m\seq0$ mode, which is similar to the bowl mode except with radial nodes. 

\cite{lyndenbell65} suggested that the mode supporting the Milky Way's integral sign warp could be a discrete mode of vertical vibration, similar to the Eulerian nutation of a coin thrown spinning into the air (a modification of the rigid-tilt mode of the disc). Later works (e.g. \citealt{hunter69, weinberg91}) discuss this $m\seq1$ mode as an outwardly propagating bending wave excited by some perturber. The exact modes are highly dependent on the halo model, with flattened halos providing support for such a warping mode. Further, the $m\seq1$ modes have not been shown to result in lasting heating of the stellar disc \citep{sellwood98}, so we will assume that owing to the lack of persistent $m\seq1$ vertical power in our simulations during the secular growth epoch that those modes are not driving buckling instabilities.

\cite{sellwood09} point out that the buckling mode depends on a variety of factors (some physical and some numerical), en route to their main point that the buckling mode can be exacerbated by stochasticity effects. Their simulations show prominent, but variable, buckling. This results in a rapid weaking of the bar-measuring $A_2/A_0$. In their Appendix B, \cite{sellwood09} examine the effect of their choice of some basic numerical parameters. If a sudden drop in $A_2/A_0$ is to be believed as a hallmark of buckling, then both the grid resolution and softening length appear to profoundly affect the buckling instability. As pointed out by \cite{sellwood06a}, the vertical resolution in a softened simulation will weaken the vertical forces and, therefore, increase the vertical oscillation period of disc particles. When the bar buckles, the in-plane motion of particles are coupled to vertical motions, meaning that a buckling mode can have a back reaction on the formation of the bar.  Whether this implicates numerical problems at large is a subject of debate.

The excitation of a vertical instability, or `buckling', commonly observed in simulations of barred galaxies, has recently been proposed to be  a generic part of the bar-formation process in disc galaxies. The buckling instability has been implicated as the primary cause of observed `peanut' bulges (see \citealt{sellwood14} for a review), though the peanut shape in barred simulations has been observed since \citet{combes81}. The original explanation of bar buckling in a simulation comes from \cite{raha91}.

They attribute the buckling to the firehose instability in the sense of \cite{toomre66} and \cite{araki85}. The vertical disturbance in \cite{raha91} is a $m\seq2$ buckling instability with a characteristic saddle shape. However, other instability explanations exist, such as the presence of a strong resonance (as argued in theory by \citealt{pfenniger98} and shown in simulations by \citealt{saha13a}) rather than a strong gradient in the velocities, as in a fire-hose instability. 

\citet{debattista06} describes what seems to be the brief ($\Delta t<100$ Myr) formation and dissipation of bending modes during the violent bar formation phase. The bending modes, when inspected by eye, appear to be a mixture of $m\seq0$ and $m\seq2$ modes, where the $m\seq2$ modes have a different pattern speed than the bar. These are long-wavelength disturbances $\lambda>2R_{\rm bar}$ that may persist, resulting in a peanut shape. In \cite{martinezvalpuesta06}, the bending modes are even more extreme, with the wavelength of the (presumably) $m\seq2$ modes increasing with each of three subsequent buckling events. They argue that the two primary explanations: the firehose instability \citep{toomre66, raha91, merritt94} and resonance heating \citep{pfenniger91} can be reconciled if buckling is merely viewed as shortening the secular timescale for particle diffusion out of the plane. \cite{sellwood09} run simulations with symmetry imposed about the midplane and find that the bar strength (measured as $A_2/A_0$) continues to grow throughout the entire simulation. This leads to an interpretation of the buckling instability as a self-regulating mechanism. Additionally, if $m\seq1$ is disabled, \cite{sellwood09} find that all simulated bars buckle violently as a result of an instability resulting from the inability of the potential to respond to a mildly lopsided distribution\footnote{We do not find this in our even--harmonic--only cusp simulation.}. Their conclusion is that it `seems unlikely that such small offsets could have such a large effect on the saturation of the buckling mode, we think it is possible that an antisymmetric mode competes'.

\cite{saha13a} studies the meridional tilt of the velocity ellipsoid in model barred galaxies, finding that the tilt reaches a peak that triggers the onset of bar buckling. They argue that the meridional tilt is a better indicator of the onset of buckling than the $\sigma_z/\sigma_r$ ratio. After the bending modes are excited, the amplitude gradually increases and drifts out to larger radii. However, \cite{saha13a} finds that a bar that grows slowly does not experience a buckling instability. They attribute the slow growth of the bar to the selection of Toomre $Q$ for the disc. Lastly, \cite{erwin16} makes a claim for observing a bar in the act of buckling, i.e., when the $m\seq1$ vertical power is largest, finding that all observed bars are consistent with having gone through a buckling phase.

Given these previous findings, one might naturally look at the strongest non-axisymmetric disturbance in the simulation as evidence for buckling. However, an inspection at the peak point of the $A_2/A_0$ in the cusp model shows that the amplitude in the vertically asymmetric terms is still weak. To identify bending modes, we isolate the vertically asymmetric terms in the basis and examine their power as a function of time. In our basis, we include three vertically asymmetric functions, $(0,9)$, $(1,10)$, and $(2,11)$\footnote{The inclusion of vertically asymmetric terms is related to the disc scaleheight and the number of radial terms included. Thus, a thick disc would naturally admit more vertically--asymmetric functions. A very thin disc would admit none in the first 12 radial orders.}. The $(2,11)$ function is shown in the lower panel of Figure~\ref{fig:disc_3d_amplitudes}.

We find mild bending modes at early times with no apparent lasting effects. The cusp and core models both exhibit mild $m\seq2$ bending after a more powerful $m\seq0$ bending mode, as indicated by the amplitude in vertically-asymmetric harmonics. The core model exhibits a peak $m\seq2$ power that is larger than that of the fiducial model. The cusp model shows a stronger $m\seq1$ harmonic subspace that is likely a consequence of the visible seiche mode. The amplitude of all the bending modes has a $|z|_{\rm max}$ of approximately $0.05a\seq0.5h$ during the $m\seq0$ dominated phase. Additionally, $|z|_{\rm max}\seq0.025a\seq0.25h$ when the $A_2/A_0$ ratio is largest in the cusp model and $|z|_{\rm max}\seq0.066a\seq0.66h$ in the core model. Our observed buckling modes are bisymmetric with respect to the disc plane, along the bar, and are confined to be within a bar radius. The modes observed in \cite{debattista06} extend past the end of the bar along the bar major axis, but are confined to be within the bar along the bar minor axis. In both our simulations, the disc does not exhibit any features that could be bending modes after $T\seq0.5$. Even at $T\seq0.5$, the disc is still quite thin, thickening only at late times after the bar has formed. We do not see any signatures of a bending mode as the disc thickens. We do observe a peanut shape for the inner region of the galaxy in both models at late times, but it does not result from a bending mode. Thus, the persistence of the peanut shape in other simulations remains a mystery. If the wavelength of the bending mode is greater than $R_{\rm bar}$, as in \citet{debattista06}, it is difficult to see how it is the parent of the peanut shape. In a key difference from our work, the length of the bar and the peanut bulge are often $>2a$ in other works, suggesting that at the very least, potential differences are at play, most likely attributable to the halo model. We defer a discussion of bulge formation and disc thickening to a later paper that explores a larger sample of models and their associated bulge formation.

\subsection{Nonlinear Modes} \label{subsec:planarmodes}

We highlight two in-plane features for their importance to bar evolution that are not explicitly captured in a linear analysis of bar evolution: the role of $m\seq1$ in bar evolution (Section~\ref{subsubsec:m1}) and short-timescale power exchange between radial functions that make up the bar (Section~\ref{subsubsec:beating}).

\subsubsection{Dipole ($m\seq1$) Modes} \label{subsubsec:m1}

\begin{figure*} \centering \includegraphics[width=6.5in]{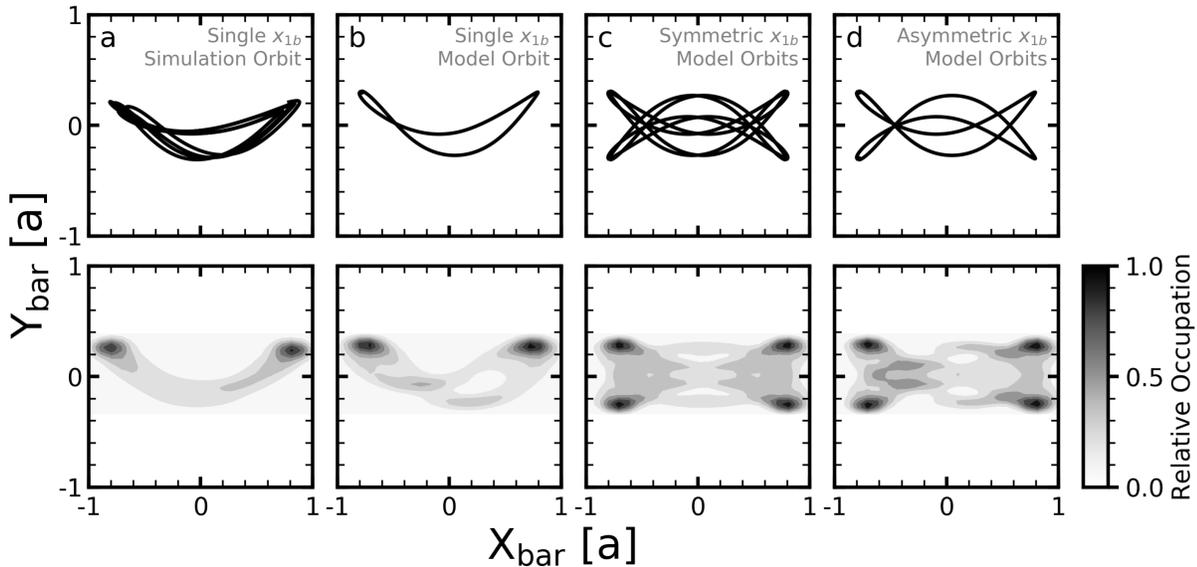} 
\caption{\label{fig:m1orbits} Illustration of the $x_{1b}$ orbits that support the increase in $m\seq1$ power at $T\seq3$ in the cusp simulation. The upper panels are orbit trajectories and the bottom panels are time-averaged densities. From left-to-right, we show a single $x_{1b}$ orbit drawn from the cusp simulation at $2.8<T<3.2$; the same simulated orbit modeled in the $T=3$ fixed potential; a collection of model orbits with each orientation equally weighted; a collection of model orbits with unequal weights of each orientation. As long as the orbits are equally represented between different orientations, no net $m\seq1$ amplitude results. However, if the orbits are unequally represented, a net $m\seq1$ amplitude results, as in the lower right panel.} \end{figure*}

The existence of $m\seq1$ modes in both the real universe and simulations has been discussed and presented many times in the literature. \citet{colin89} demonstrated that offset bars would have a deforming effect on the morphology of galaxies, re-locating the Lagrange points that are crucial for parenting stable orbits just outside the bar radius. However, our post-simulation orbit analysis forces the bar and halo wake to have the same pattern speed for several rotation periods, which is not true in self-consistent simulations\footnote{The pattern speed of the $m\seq1$ and $m\seq2$ components do in fact reach an equilibrium in our fiducial simulation, as we discuss below.}. \cite{athanassoula96} reproduced one-armed spiral morphology in simulations and noted the importance of the impact position with respect to the bar. In the following year, \cite{athanassoula97} found that the displacements of the centers are accompanied by changes in the bar pattern speed and bar size.

On the observational side, \cite{zaritsky13} studied $m\seq1$ distortions in 167 galaxies from the S$^4$G sample of nearby galaxies to determine the origin of lopsidedness in galaxies. While $m\seq1$ largely increased with radius, the $m\seq1$ strength was not related to the presence or absence of a bar, or bar strength if a bar were present. \cite{saha14} examined angular momentum transport in lopsided galaxies through the paradigm of \cite{lynden72}. An extreme example of $m\seq1$ power might be the Large Magellanic Cloud (LMC) \citep{pardy16}, which exhibits both a one-armed spiral and a bar offset from the dynamical center of mass. The Milky Way and Small Magellanic Cloud have long been implicated in the formation of structure in  the LMC, and it stands to reason then that the MW or SMC would be the cause of the $m\seq1$ disturbance. \citet{pardy16} modeled the LMC-MW interaction and found that the stellar disc of the LMC was shifted away from the dynamical center of the galaxy, rather than the bar itself being shifted to being off-center. 

Despite the clear contribution of the bar to the $m\seq1$ harmonics and a non-zero amplitude during our simulations, the orbit analysis in \citetalias{petersen18a} ignored the existence of the $m\seq1$ harmonics and still created orbital structures that matched the structure observed in the self-consistent simulations (both cusp and core). Therefore, we are left wondering whether the $m\seq1$ effect is a true physical effect, or merely excited by noise in the simulations with little true effect on the evolution. Both could be true, in the sense that noise in the simulation may be physical in origin rather than numerical, where the $m\seq1$ harmonics are required to adequately resolve the stochastic excitation of $m=1$. Evidence bolstering this conclusion comes from the cusp simulation run using only even azimuthal harmonics. The bar that forms in the even--harmonic--only cusp simulation is appreciably different from that in the full simulation: it reaches only half the maximum amplitude of the cusp bar, and is more compact. Given that the $m\seq1$ amplitude during the assembly phase is not attributable to the bar itself, the initial $m\seq1$ likely relates to a readjustment of the disc that enables the bar to grow further.

The excitation of $m\seq1$ at early times clearly relates to the formation of the bar, and damps before $T\seq1$ in both models (upper panel of Figure~\ref{fig:bar_fraction}). At $T\seq3$ in the cusp simulation, we see oscillations in both $m\seq2$ and $m\seq1$ power that appear to mirror each other, which probably owes to power exchange: `mode-coupling', which we labeled as `interaction' in Figure~\ref{fig:cusp_phases}. The mode locking appears to be a feature of the specific cusp model and is not observed in the core model, but there is likely a class of models with a combined disc and halo that have a higher $m\seq1$ pattern speed (cf. bottom panel of Figure~\ref{fig:cusp_phases}). A slightly different model may not result in mode locking. In this sense, it is not coincidental that the asymptotic value of $\Omega_2$ is half that of $\Omega_1$. 

Such mode coupling is a nonlinear process. Simulations may be the best path forward to understanding its dynamical implications. In the cusp simulation, we conclude that harmonic coupling results from the bar attempting to transfer angular momentum into any reservoir available, and it finds the $m\seq1$ harmonic as an extra dynamical degree of freedom, even while the disc transfers $\Lz$ to the combined $m\seq1$ and $m\seq2$ system.

Whether the two patterns may phase lock and still exchange $\Lz$ is still an open question that requires more study. An analysis of the torque induced during the mode-locking phase at $T\seq3$ of the cusp simulation \citepalias{petersen18b} reveals that the torque on the bar by both the outer disc and halo decreases with an increase in $m\seq1$ amplitude. We conclude that the $m\seq1$ harmonic is being torqued by the outer disc as it attempts to transport angular momentum inward, resulting in an increased $m\seq1$ amplitude but a decreased torque on the bar. 

Using an orbit analysis of the bar \citepalias{petersen18a}, we find that the $m\seq1$ bar feature is supported by the phase coherence of the asymmetric $x_{1b}$ orbits. The asymmetric $x_{1b}$ orbits form their own self-gravitating feature as the remaining symmetry axis is broken in the strong bar, resulting in an increased $m=1$ amplitude. The strong bar pumps energy into the natural asymmetry of the $x_{1b}$ orbits if the bar fluctuates in its centroid, resulting in a parametric resonance. An illustration of the mechanism that generates $m\seq1$ amplitude from $x_{1b}$ orbits is shown in Figure~\ref{fig:m1orbits}. In \citetalias{petersen18a}, we found that $x_{1b}$ orbits naturally arise during the growth phase of the bar, and are present in some fraction during the steady-state phase (cf. Figure~\ref{fig:cusp_phases}).  At time $T<2.8$, the fraction of orbits that are in the various possible orientations of $x_{1b}$ orbits are equal. During the mode-locking event, the fraction becomes heavily weighted ($\approx0.75$) toward the orbits that preferentially loop on one side of the bar. This symmetry-breaking is enough to support the $m=1$ amplitude.  In principle, any unequal representation of $x_{1b}$ orientations can create $m\seq1$ power. The determination of orbits reinforcing the $m\seq1$ amplitude and the reduction of torque on the bar to zero supports our conclusion that the $m\seq1$ amplitude  is a real dynamical effect with evolutionary importance.

If systems can induce mode locking with low-level integer commensurabilities, the exchange of power will have wide-ranging implications for the dynamics. The frequency range for the $m\seq1$ pattern is narrow \citep[e.g.][]{weinberg94} and the bar pattern speed changes by a factor of 2-3 during its evolution (see Figures~\ref{fig:cusp_phases} and \ref{fig:core_phases}). Therefore, it seems likely that the frequency of $m\seq2$ (bar) will be commensurate with that of $m\seq1$ (seiche) at some point during its evolution. For a system with weakly damped $m\seq1$ modes, an interaction with $m\seq2$ would be expected. Unfortunately, the method used in \cite{weinberg94} to determine the frequency of the allowed $m\seq1$ modes, calculating the dispersion relation for the halo-disc system, only applies to weak perturbations and not to a strongly barred galaxy like we have here. We, therefore, must resort to a larger suite of barred galaxy models (Petersen et al. 2019d) to predict the importance of this mode--locking in Nature.

We find that the coupling between the $m\seq1$ and $m\seq2$ harmonic subspaces exists well above the Poisson noise level by a factor of 100 in amplitude. We also track the center-of-mass for both the disc and halo expansion, finding that the offset of the two is never larger than 5 per cent of a disc scalelength. Therefore, force errors caused by basis incompleteness are unlikely to be an issue, thereby alleviating the concern that mode locking is numerical in origin.

\subsubsection{Power exchange between radial functions} \label{subsubsec:beating}

Some responses are identified in the basis coefficients through power exchange between two different harmonic subspaces, which may be either azimuthal (as above between $m\seq1$ and $m\seq2$), or between radial orders in the same azimuthal subspace (Figure~\ref{fig:radial_order}). We observe that the coefficients exchange power on a bar rotation timescale. Power exchange indicates nonlinear behavior that is not easily captured by other methods, so we study the possible dynamical contributions of power exchange to our models here. We specifically ask whether power exchange between different coefficients affects bar evolution.

As an example, we consider the $(2,1)$ and $(2,2)$ terms in the cusp model (the lower left panel of Figure~\ref{fig:radial_order}). We see that the amplitude of a given $m\seq2$ radial order can change by 20 per cent in opposing directions as a result of high-frequency variation in the coefficients. However, the overall amplitude of the entire $m\seq2$ harmonic does not exhibit any higher-order variation (cf. $A_2/A_0$ in Figure~\ref{fig:cusp_phases}). One may naturally question whether this rapid variation has an effect on the resultant potential that would bias a detailed analysis of the orbit structure. For example, selecting a snapshot of the potential at $T\seq1.75$ versus $T\seq1.8$, less than a bar rotation later, would result in a different $n$ spectrum of amplitudes. However, the radial harmonics appear to be matched in phase, suggesting that the amplitude variation relates to changing structure in the bar rather than to the presence of mode--locking or beating.  Ideally, one would like to study the spatial correlations between basis features and the coefficient time series, but such a study is beyond the scope of this paper. We find that small-scale power exchange does not affect our method to determine and interpret the orbital structure on short timescales, such that one may confidently analyze the evolution of the bar over the $\Delta T\approx1$ phases we identify.

A second example of higher-frequency power exchange illustrates that power exchange is unlikely to be a numerical artifact. In the $m\seq2$ amplitude at $T>3$ in the core simulation, as shown in the lower right panel of Figure~\ref{fig:radial_order}, the amplitude of variations suddenly shifts from of order 5 per cent to of order 1 per cent. The frequency of the power exchange changes to a higher order multiple of the bar period as the pattern speed decreases, as expected, suggesting that this interaction is dynamical in nature and not a numerical artifact. We also find that the bar pattern is the main driver of the power exchange with a 1:1 frequency ratio, suggesting that the bar beats with the outer disc. The semi-periodic variations in the bottom panels of Figure~\ref{fig:bar_fraction}, where the fraction of amplitude in the bar trades off with that in the untrapped disc on roughly a bar period in both simulations, also leads us to conclude that the bar beats against the outer disc. 

Thus, the pattern of power exchange in the coefficients suggests that: (1) interpretations of bar structure from linear, fixed potential analysis \citepalias[e.g.][]{petersen18a} are not strongly affected by any power exchange between radial orders within a harmonic order, and (2) power exchange between radial orders corresponds to resolvable dynamical processes, not to artifacts in the basis.

\section{Observational Diagnostics} \label{sec:observations}

We discuss two different observational diagnostics for bars. In Section~\ref{subsec:dressing}, we describe orbits that bias ellipse fits of bars and demonstrate how ellipses may overestimate the length of the bar. In Section~\ref{subsec:kinematics}, we present a kinematic method to determine the maximal length of the `backbone' $x_1$ orbits in real galaxies.

\subsection{Dressing Orbits} \label{subsec:dressing}

\begin{figure*} \centering \includegraphics[width=5.5in]{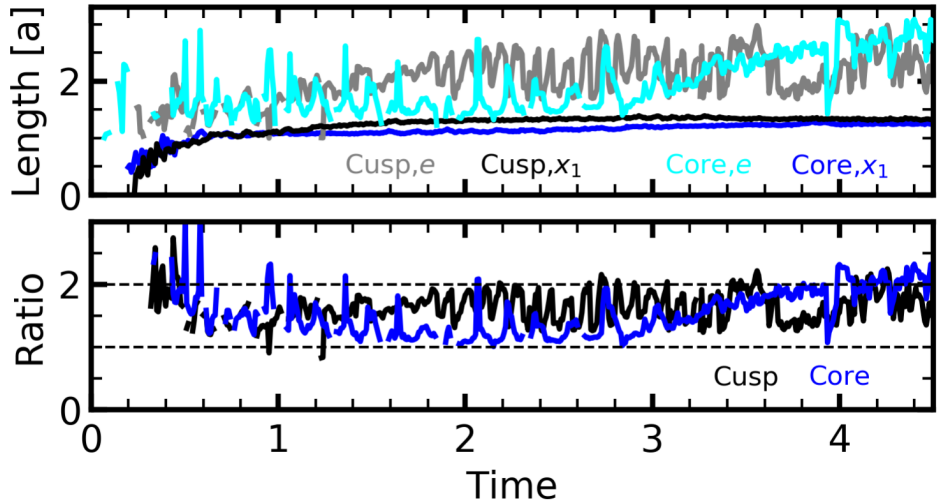} 
\caption{\label{fig:ellipse_fit} Upper panel: The length of the bar in disc scalelengths, measured in both simulations (cusp and core), using two different techniques: maximal $x_1$ extent and ellipse fits, versus time. The cusp simulation ellipse-fit-derived length is shown in gray and its $x_1$-derived length is in black. The core simulation ellipse-fit-derived length is shown in cyan and its $x_1$-derived length is in blue. Lower panel: The ratio of the ellipse-fit-derived length to the $x_1$--derived length versus simulation time for the cusp (black) and core (blue) simulations. } \end{figure*}

\begin{figure} \centering \includegraphics[width=3.5in]{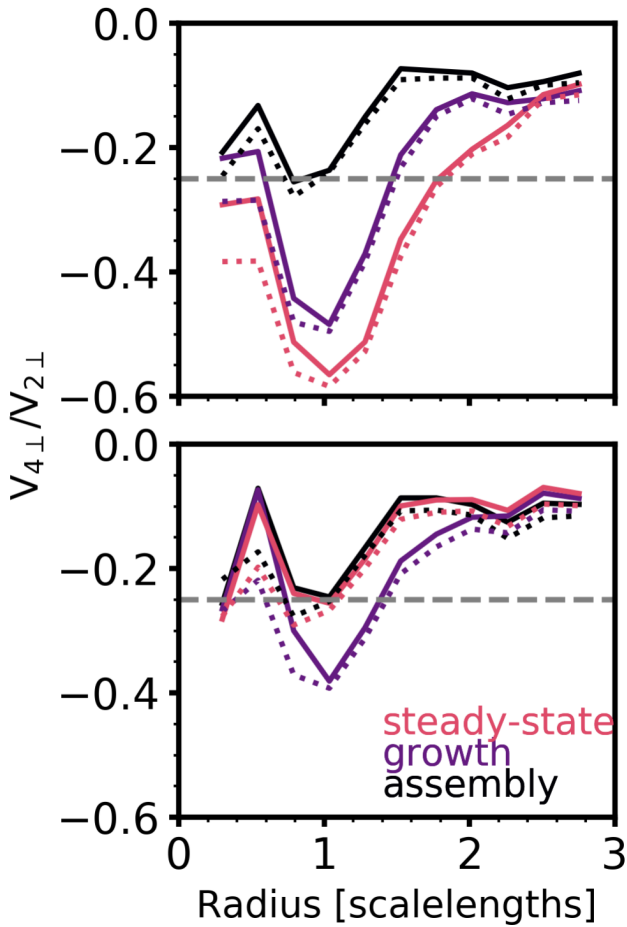} 
\caption{\label{fig:a4a2diagnostic} $v_{y\perp}/v_{2\perp}$ as a function of radius in annular bins, computed from the magnitude of the velocity tangential to the bar, $v_\perp$. The solid lines are computed from images degraded to $\delta v=0.05$ resolution, and the dotted lines are from images degraded to $\delta v=0.02$ resolution. The radius minimizing $v_{4\perp}/v_{2\perp}$ corresponds to the approximate maximal extent of the $x_1$ family of orbits present in the model. The value of the minimum indicates the strength of the measurement. The dashed line at $v_{y4}/v_{y2}\seq-0.25$ is our empirically-determined threshold for a trustworthy measurement. The upper panel is for the cusp simulation and the lower panel the core simulation.} \end{figure}

\begin{figure*} \centering \includegraphics[width=6.5in]{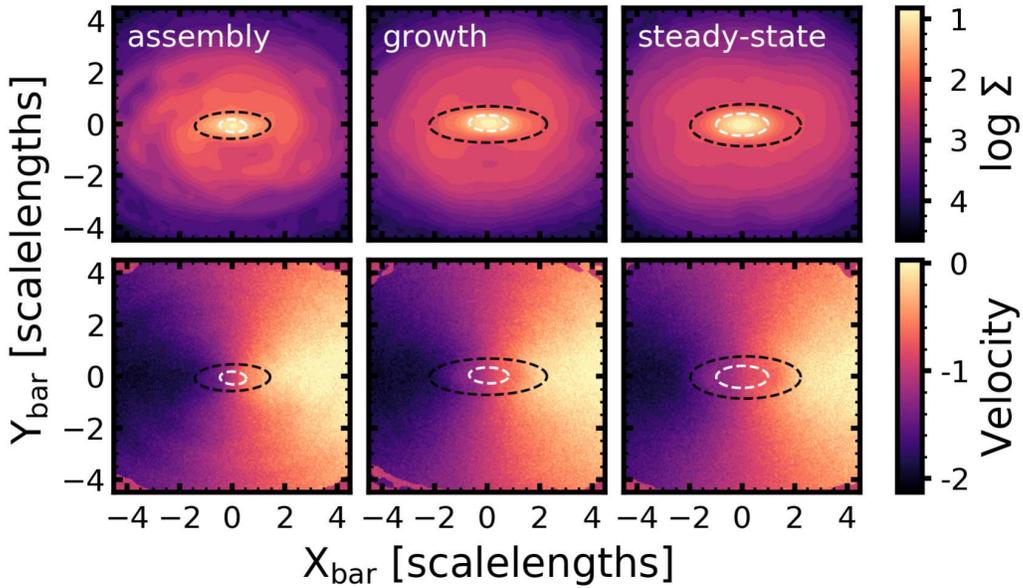} 
\caption{\label{fig:cusp_velocity} Upper panels: Log surface density, in normalized units, for three phases in the cusp simulation, where the simulation has been rotated so it is viewed at a 45$^\circ$ angle: assembly, growth, and steady-state. Lower panels: Velocity field in the direction tangential to the bar, for the three phases in the upper panels, degraded to a velocity resolution of $0.1v$ (10 $\kms$). The white dashed ellipses show the maximum extent of the trapped $x_1$ orbits, which coincides with the dimple in the velocity field of the growth and steady-state phases, as calculated from $v_{4\perp}/v_{2\perp}$ for the velocity field tangential to the bar (see text). The black dashed ellipses show the best-fit ellipse for the bar from the surface density plot alone.} \end{figure*}

\begin{figure*} \centering \includegraphics[width=6.5in]{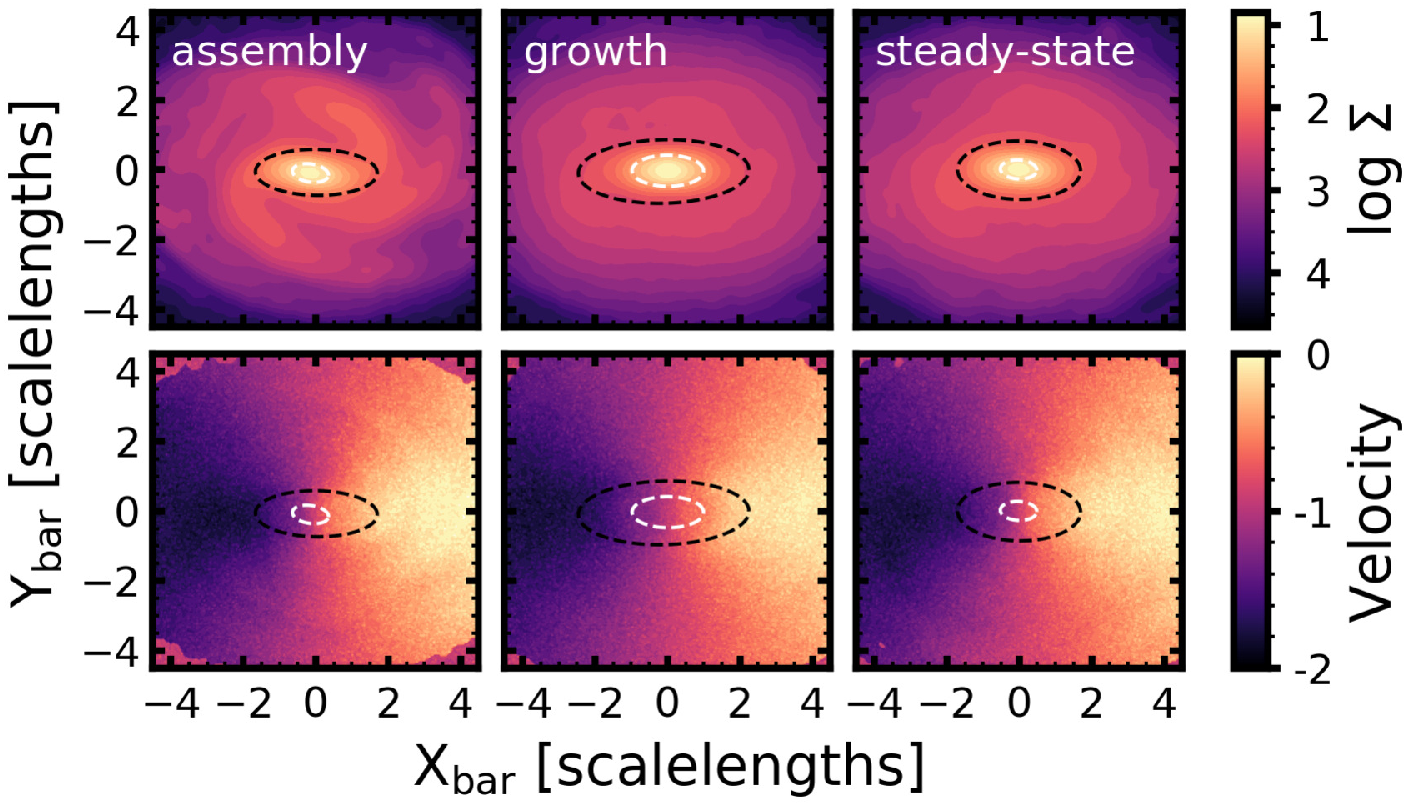} 
\caption{\label{fig:core_velocity} The same as Figure~\protect{\ref{fig:cusp_velocity}} but for the core simulation. The three columns correspond to the phases of evolution, preserving the order as in Figure~\protect{\ref{fig:cusp_velocity}}, though in time, the steady-state comes before the growth phase in the core simulation.} \end{figure*}

We discuss relevant observational diagnostics and the possible pitfalls inherent in attempting to measure bars from surface density measurements. We describe how the standard ellipse--fitting approach gives biased quantities when compared to the maximal extent of truly trapped bar orbits in the $x_1$ `backbone' family. In practice, the maximal extent of the $x_1$ family is never truly reached, as demonstrated in the orbit analysis of \citetalias{petersen18a}, so methods that parameterize the length of the bar as the maximal theoretical $x_1$ extent may also overestimate the actual bar length \citep{martinezvalpuesta06}.

We refer to orbits that are not trapped by the bar, but are still in the physical vicitinity of and affected by the bar as `dressing' orbits. Dressing orbits confuse the normal measurement metrics, in particular ellipse fits and Fourier-derived strengths. A standard ellipse fit to the bar may overestimate the mass and length of the bar by a factor of two! Typically, the length of the bar will be overestimated by 50 per cent. One needs an accurate length for the true bar orbits to observationally determine the pattern speed (see e.g. \citealt{perez12}). The dimensionless parameter $\mathcal{R} \seq R_{\rm CR}/a_{\rm bar}$, where $R_{\rm CR}$ is the corotation radius and $a_{\rm bar}$ is the semi-major axis of the bar as given in \citet{binney08}, denotes the `slowness' of the pattern speed. As discussed in \citet{binney08}, this parameter can be hard to measure in real galaxies for two reasons: (1) the bar does not have a sharp end, and (2) corotation does not have a clear definition for strong non-axisymmetric disturbances, e.g. a strong bar. However, several studies have attempted to measure either the pattern speed or the dimensionless parameter to characterize the bar. Thus, an overestimate for the true dynamical length of the bar overestimates the pattern speed, sometimes significantly.

Assuming a constant mass--to--light ratio, we apply the standard ellipse-fitting analysis to our simulations. We compute the face-on surface density at a resolution of $0.025a$, which for a MW--like galaxy corresponds to 75 pc\footnote{In practice, the ellipse fits do not turn out to be highly dependent on the resolution, introducing approximately a 10 per cent error.}. We measure the length of the bar using a standard method: determine best-fit ellipses at many different surface densities and assign the bar length to the semi-major axis value, $a$, where the ellipticity drops below a certain threshold or has a discontinuity. Here, we choose the semi-major axis where the ellipticity $e\equiv1-\frac{b}{a}$, and $b$ is the semi-minor axis value, first drops below 0.5. In practice, this is also the same location as the discontinuity in $e$ that corresponds to the transition between bar--dominated contours and disc--dominated contours. We also have a dynamically-informed metric for the length of the bar: the maximum extent of the $x_1$ family \citepalias{petersen18a}. Figure~\ref{fig:ellipse_fit} shows a comparison of the maximum $x_1$ extent versus ellipse--fit derived bar lengths. In the upper panel, we show the measured bar lengths for the cusp and core simulation measured using both techniques. The ellipse fit length in the cusp simulation (gray line) grows steadily at $T<2.5$. This roughly mirrors the trapped fraction growth, but when compared to the maximal extent of the $x_1$ orbits (black), we see that this is an extreme overestimate for the length of the trapped component. The periodicity in the ellipse measurements results from the outer disc $m\seq2$ disturbances coincidentally aligning with the bar. At early times when the bar is forming, this can result in variations of nearly a factor of two. Even at later times in the cusp simulation, the variation in ellipse--fit length is 25 per cent over short ($\delta T\seq0.1$) timescales owing to the $m\seq2$ alignment.

For the core simulation, the ellipse measurements oscillate during the assembly epoch (cyan in Figure~\ref{fig:ellipse_fit}), as spiral arms align and anti-align with the bar on bar-period timescales. As the simulation progresses, the variations from the outer disc $m\seq2$ features weaken, and the ellipse--fit appears to largely agree with the maximal $x_1$ (blue for the core simulation). However, as the core simulation enters the growth phase near $T\seq3$, the two measurements begin to diverge. After some growth in the trapped fraction (cf. Figure~\ref{fig:core_phases}), the shallow surface density profile at the end of the bar conspires with the lengthening bar to find new ellipse contours and the ellipse--fit length increases rapidly. While this scenario may not occur in every model, it is not an artifact of tuning as we adopted a standard implementation of ellipse fitting procedues and parameters. There is little reason not to suspect a similar behavior in real galaxies. We conclude that the length of a bar as measured from ellipse fits should not be interpreted as an age indicator.

The lower panel of Figure~\ref{fig:ellipse_fit} summarizes the overall results of our comparison. We plot the ratio of the ellipse--fit length to the maximal $x_1$ length for the cusp (black) and core (blue) simulations. The ellipse--fit length is a large overestimate for the length of the $x_1$ orbits at all times in the cusp simulation, typically by a factor of 1.5, except during the assembly phase when the overestimate is a factor of two. In the core simulation, the ellipse--fit length is a better estimate for the maximum extent of the $x_1$ orbits during the steady-state phase near $T\seq2$, but overestimates the length during assembly  by a factor of 1.5, and overestimates the length at late times ($T>3.4$) by a factor of two after the bar stars to grow by trapping. Taken together, the cusp and core simulations reveal the ambiguity in ellipse--fit determinations of bars. Ellipse fits should be taken as a measure of the total mass distribution of the galaxy, not as the mass directly associated with the bar itself.

\subsection{Kinematic Signatures}\label{subsec:kinematics}

We describe a kinematic diagnostic that can measure the maximal extent of $x_1$ orbits using current and future generation integral field units (IFUs). The technique works by exploiting the velocity tangential to the bar axis, which for trapped $x_{1}$ bar orbits will be low compared to disc orbits. The signal will be largest where the discrepancy between bar orbits and disc orbits is largest, i.e. at the `four corners' of the bar. This suggests that a kinematic metric using four--fold $m\seq4$ symmetry will reveal the largest differences between bar and disc velocities. The difference between the velocities of bar and disc orbits will be negative at the corners of the bar and so we expect the $m\seq4$ velocity moment tangential to the bar, $v_{4\perp}$, to be appreciably negative. As the bar slows, we expect this quantity to become even more extreme since the velocity between the untrapped disc orbits and the bar pattern speed becomes more discrepant \citepalias{petersen18a}. 

To test the significance of the signal relative to that of the bar, we compare the $m\seq2$ and $m\seq4$ velocity moments, $v_{4\perp}$ to $v_{2\perp}$. The procedure to observationally determine the maximal extent of $x_1$ orbits is as follows:
\begin{enumerate}
\item Compute the {\it magnitude} of the velocity perpendicular to the bar, $v_\perp$.
\item Compute the $m\seq2$ and $m\seq4$ Fourier velocity components as a function of radius, $v_{2\perp}(r_j)$ and $v_{4\perp}(r_j)$, where the $\{r_j\}$ are annular radii. One must take care not to reduce the S/N by choosing annuli that are too narrow relative to the spatial resolution. We suggest a minimum annular radius of $\delta r \seq 10\delta x$, where $\delta x$ is the pixel scale.
\item Locate the radius for the ${\rm min}\left\{v_{4\perp}(r_j)/v_{2\perp}(r_j)\right\}_j$. As long as ${\rm min}\left\{v_{4\perp}(r_j)/v_{2\perp}(r_j)\right\}_j<-0.25$ this method is reliable and the minimum approximately equals the maximal extent of the $x_1$ orbits.
\end{enumerate}

To use this metric, one requires high spatial and velocity resolution, coupled with a modest inclination. As a general guideline, the spatial resolution required to determine the $v_{y\perp}/v_{2\perp}$ metric is $\delta x\seq0.05a$, where $a$ is the disc scalelength, for a galaxy at an inclination of 45$^\circ$. The velocity resolution required is $0.05v_{\rm max}$, where $v_{\rm max}$ is the maximum circular velocity. For a MW-like galaxy, this translates to $\approx$10$\kms$ velocity resolution and 150 pc spatial resolution. We determine the threshold ${\rm min}\left\{v_{4\perp}(r_j)/v_{2\perp}(r_j)\right\}_j<-0.25$ empirically by applying the method to simulated galaxies during the assembly phase, for which we do not expect the velocity field to recover the bar feature, and compute the ${\rm min}\left\{v_{4\perp}(r_j)/v_{2\perp}(r_j)\right\}_j$ value obtained spuriously, finding this value to be approximately -0.25. Current image--slicing instruments such as MUSE should be able to detect bar length using this method. For example, the data published in \cite{gadotti15} featured 12 pc spatial resolution with $\approx$10$\kms$ velocity resolution in the nearby barred galaxy NGC 4371 ($d\seq16.9$ Mpc) using MUSE. While the spectral resolution is at the limit of what is needed to detect the $v_{y\perp}/v_{2\perp}$ effect, the superior spatial resolution provides an excellent opportunity to look for velocity features directly attributable to a particular orbit family.

 In Figure~\ref{fig:a4a2diagnostic}, we show results using the method, for both the cusp (upper panel) and core simulations (lower panel). As expected, both simulations show negative $v_{y\perp}/v_{2\perp}$ values, driven by the effect at the corners of the bar. The solid lines are computed from velocity images degraded to $0.05v_{\rm max}$ in annular bins that are $0.15a$ in width, for a galaxy inclined\footnote{In practice, the inclination of the galaxy merely weakens the signal; if the minima satisfies ${\rm min}\left\{v_{4\perp}(r_j)/v_{2\perp}(r_j)\right\}_j<-0.25$, the result is trustworthy.} at 45$^\circ$ and with a bar position angle of 0$^\circ$. We compute the dotted lines with velocity images degraded to $0.02v_{\rm max}$. We have tested the metric for a range of inclination angles, position angles, velocity resolutions, and spatial resolutions to develop the observational guidelines presented above. In the cusp simulation (upper panel), the signal is very strong during the growth and steady-state phases, with no discernable signature in the assembly phase. One expects a low signal in the assembly phase since the kinematic feature results from trapped, evolved orbits that develop at the start of the growth phase. In the core simulation (lower panel), the signal is particularly strong during the growth phase, although we are able to tease out accurate lengths in the assembly and steady-state phases that are marginally significant, owing to the smaller fraction of $x_1$ orbits. Even in the case of marginally significant detections, the  $v_{y\perp}/v_{2\perp}$ method will result in a more accurate bar length compared to ellipse fits.

Higher velocity and spatial resolution, not feasible using current instruments, makes a modest difference, particularly in the case of the core simulation. Comparison with maximal $x_1$ extents \citepalias[computed in ][as shown in Figure~\ref{fig:ellipse_fit}]{petersen18b} shows that the minima of $v_{y\perp}/v_{2\perp}$ is within 10 per cent of the maximal $x_1$ orbit radius, making this technique a powerful descriminator of the dynamically-relevant maximal $x_1$ orbit.

In Figures~\ref{fig:cusp_velocity} and \ref{fig:core_velocity}, we compare our velocity moment method with standard ellipse fits. The upper panels of Figure~\ref{fig:cusp_velocity} show the surface density during the three identified phases of bar evolution, and the lower panels show the $y$-component of the velocity (perpendicular to the bar) for each of the three phases. For illustrative purposes of this method in practice, the galaxies have been inclined to $i=30^\circ$ relative to the page. The velocity resolution has been degraded to a velocity resolution of $0.1v$ (10$\kms$) by injecting random noise into the measured velocity field. In each phase, we plot the standard technique (ellipticity drop) best-fit ellipse in dashed black. The ellipse that corresponds to the minimum of $v_{y\perp}/v_{2\perp}$ is shown in dashed white. As shown in Figure~\ref{fig:a4a2diagnostic}, while we may compute a minima in the $v_{y\perp}/v_{2\perp}$ value for the assembly phase, it is too small to say anything about the orbit structure with certainty\footnote{With the omniscience provided by simulations and the true calculation of the maximal $x_1$ orbits, we see that the $x_1$ track is not yet fully formed (the apsis precession which assembles the bar is an ongoing process), and thus the technique will not, by definition, be informative.}. However, during the growth and steady-state phases, strong signals are observed in $A_4/A_2$ which show the maximal extent of the $x_1$ orbit family. As described in Section~\ref{subsec:dressing}, the discrepancy between ellipse-fit lengths and maximal $x_1$ orbits can be significant. In the core simulation (Figure~\ref{fig:core_velocity}) we see much the same effect as in the cusp simulation, and we are able to calculate a relevant $x_1$ bar length during the assembly phase owing to the rapid construction of the $x_1$ family (cf. Figure~\ref{fig:core_phases}). As in Figure~\ref{fig:ellipse_fit}, the best-fit ellipses and the maximal $x_1$ extent are appreciably different.

This technique is a simple, albeit an observationally expensive way to search for the dominant barred galaxy orbit, the $x_1$ family. Determining the maximal extent of the $x_1$ family is the first step for determining a dynamically-relevant length of the bar, and a more accurate measure of the trapped fraction in galaxies.

\section{Conclusions} \label{sec:conclusion}

We studied two MW-like barred disc models embedded in dark matter halos using harmonic analysis enabled by the biorthogonal basis intrinsic to the BFE potential solver {\sc exp} that we use for our gravitational potential solver. The initial disc profiles are identical while one halo profile is cuspy and one is cored. We describe the two barred galaxy models in terms of azimuthal and radial harmonics, correlating previously obained features of orbits families and evolutionary phases with radial and azimuthal harmonics. 

The main results are as follows:
\begin{enumerate}
\item Decomposing barred galaxy models using BFE provides a qualitative description of evolutionary phases for barred galaxies, and provides computationally inexpensive diagnostic power for the underlying structure of model evolution. 
\item The bar is responsible for the vast majority of the $m\seq2$ and $m\seq4$ amplitude in the simulation ($>80$ per cent), rather than the observed spiral arms. Additionally, the bar is responsible for exciting a significant amount of the $m\seq1$ amplitude in the cusp model ($\approx75$ per cent), and somewhat less in the core model ($\approx60$ per cent), owing to the lower central density of the cored halo.
\item We analyze the successes and failures of observational techniques meant to characterize bars. We compare harmonic analysis measures to the trapped fraction of the bar. We find that while the harmonic analysis reproduces the qualitative evolutionary trends after the bar is established, harmonic function analysis does not elucidate the bar assembly phase, and thus cannot be used to understand the formation of bars in a straightforward manner. 
\item Observational techniques currently used on both real galaxies and simulated galaxies are not measuring true dynamical quantities. We perform ellipse fits on our simulations, finding that typical ellipse--fit techniques systematically overestimate the maximum radial extent of the trapped bar orbits. We describe why ellipses will overestimate the length, and therefore mass, of the bar. Ellipse-fit methods do not accurately represent the radial extent nor mass of orbits trapped in the bar, but ellipse-fit methods may reproduce the trends in evolution seen in simulations  {\it after} the bar has fully formed.
\item We show that the remaining non-bar percentage of nonaxisymmetric amplitude is responsible for significantly biasing observations of barred galaxies. `Dressing' orbits, those which are spatially coincident with the visual bar feature but are untrapped, can appreciably change the perceived and measured strength of the bar. 
\item IFU stellar velocity data enables locating orbits trapped in the bar feature using a simple Fourier-based velocity diagnostic, $v_{y\perp}/v_{2\perp}$. The bar-length bias caused by dressing orbits can be mitigated through the inclusion of velocity data.
\item The dipole $m\seq1$ response is a consequence of reaching a steady-state equilibrium in the cusp simulation and is related to both the orbital structure \citepalias[from][]{petersen18a} and angular momentum transfer or torque \citepalias[as in][]{petersen18b}. The signature of this event is harmonic mode coupling between the $m\seq1$ and $m\seq2$ azimuthal harmonics and can greatly affect late-time bar evolution.
\item We do not find any evidence for a buckling instability in our simulations. We explain the physical reasons for why we do not find the lack of a vertical instability to be a surprise---primarily the slow growth of the bar. The models still grow bulges, suggesting that bars do not need to buckle to produce observed boxy or peanut bulges.
\item Fixed potential analysis may be used, even in the presence of nonlinear effects such as harmonic coupling or power exchange between radial orders, to reasonably characterize the orbit structure. A fixed potential analysis is unable to fully describe the evolution of bar phases on its own: one needs BFE-based harmonic analysis to resolve nonlinear evolutionary scenarios.
\end{enumerate}

The `summary' nature of BFE, where the three-dimensional potential of the total disc-halo system is described by eigenfunctions and $\approx1000$ coefficients in azimuthal and radial harmonics ($m$ and $n$ respectively), enables fixed potential studies as in the companion work \citetalias{petersen18a}. Our BFE methodology enables new avenues for studying the evolution of bars impossible using simple Fourier analyses. Fourier decomposition methods do not give an accurate physical description of the influence of the bar, as characterized by orbit classification \citepalias{petersen18a}. 

In another companion work, we have used the BFE method to show the physical influence of the bar on the evolution of the system during the bar phases \citepalias{petersen18b}. Future work will extend the harmonic analysis techniques to a larger suite of model initial conditions to answer whether the phases of bar evolution seen in these simulations are ubiquitous.  We hope to invert the inquiry and make predictions about the evolution of galaxies from harmonic function analysis alone, or to even specify evolution in a disc galaxy model using a defined set of known harmonics. It is clear that the true power to understand model galaxy evolution comes from a hybrid suite of analyses, including trapping analysis \citepalias{petersen18a}, torque analysis \citepalias{petersen18b}, and harmonic function analysis (this paper).

\bibliography{PetersenMS}

\end{document}